\documentclass[journal]{IEEEtran}
\IEEEoverridecommandlockouts
\usepackage{amsmath}
\usepackage{graphicx}
\usepackage{amssymb}
\usepackage{cases}
\usepackage[ruled,vlined,lined,ruled,linesnumbered]{algorithm2e}
\usepackage{cite}
\usepackage{fancyhdr}
\usepackage{relsize}
\usepackage{float}
\usepackage{longtable}
\usepackage{nomencl}

\begin{document}
%\chead{IEEE International Workshop on Smart Grid Communications and Networks, Houston, USA, 2011.}
\title{Using Covert Topological Information for Defense Against Malicious Attacks on DC State Estimation}
\author{Suzhi~Bi, ~\IEEEmembership{Member,~IEEE} and Ying Jun (Angela) Zhang,~\IEEEmembership{Senior Member,~IEEE}
        \thanks{This work was supported in part by the National Natural Science Foundation of China (Project number 61201261), the National Basic Research Program (973 program Program number 61101132) and the Competitive Earmarked Research Grant (Project Number $419509$) established under the University Grant Committee of Hong Kong.}
        \thanks{S.~Bi is with the Department of Electrical and Computer Engineering, National University of Singapore, Singapore 119077. (Email: bsz@nus.edu.sg)}
        \thanks{Y.~J.~Zhang is with the Department of Information Engineering, The Chinese University of Hong Kong, Shatin, New Territories, Hong Kong, and Shenzhen Research Institute, The Chinese University of Hong Kong, Shenzhen, China. (Email: yjzhang@ie.cuhk.edu.hk)}}

\maketitle

\begin{abstract}
Accurate state estimation is of paramount importance to maintain the power system operating in a secure and efficient state. The recently identified coordinated data injection attacks to meter measurements can bypass the current security system and introduce errors to the state estimates. The conventional wisdom to mitigate such attacks is by securing meter measurements to evade malicious injections. In this paper, we provide a novel alternative to defend against false-data injection attacks using covert power network topological information. By keeping the exact reactance of a set of transmission lines from attackers, no false data injection attack can be launched to compromise any set of state variables. We first investigate from the attackers' perspective the necessary condition to perform injection attack. Based on the arguments, we characterize the optimal protection problem, which protects the state variables with minimum cost, as a well-studied Steiner tree problem in a graph. Besides, we also propose a mixed defending strategy that jointly considers the use of covert topological information and secure meter measurements when either method alone is costly or unable to achieve the protection objective. A mixed integer linear programming (MILP) formulation is introduced to obtain the optimal mixed defending strategy. To tackle the \emph{NP-hardness} of the problem, a tree pruning-based heuristic is further presented to produce an approximate solution in polynomial time. The advantageous performance of the proposed defending mechanisms is verified in IEEE standard power system testcases.
\end{abstract}


\section{Introduction}
\subsection{Motivations and summary of contributions}
\IEEEPARstart{S}{tate} estimation is a major component in the Energy Management System (EMS) of electrical power grids. The state estimator receives the raw meter measurements fed by the SCADA (Supervisory Control and Data Acquisition) system, filters the incorrect data and derives the optimal estimate of the power system operating states \cite{2004:Abur}. The state estimates will then be passed on to the other EMS application functions such as the contingency analysis and optimal power flow, to control and optimize the system performance. Accurate state estimation is critical to maintain the system to operate in a secure and efficient state. However, the integrity of state estimation is under mounting threat as we gradually transform the current electricity infrastructures to future smart power grids. The system is more open to the outside networks from the extensive use of internet-based protocols in the communication system. In particular, enterprise networks and even individual users are allowed to connect to the power network information infrastructure to facilitate data sharing \cite{2008:Ten}. With these entry points introduced to the power system, potential complex and collaborating malicious attacks are brought in as well. Liu \emph{et al}.\cite{2009:Liu} showed that a false-data injection attack that alters a subset of selected meter measurements could circumvent bad data detection (BDD) in today's SCADA system and introduce arbitrary errors to state estimates without being detected. Essentially, the altered meter measurements are structured to be consistent with the physical power flow constraints. Such an attack is referred to as an \emph{undetectable} false-data injection attack and has attracted increasing research interest in the power system cyber-security \cite{2013:Giani,2010:Dan,2012:Cui}. A recent experiment in \cite{2011:Teixeira} demonstrates that the attack can cause a state-of-the-art EMS/SCADA state estimator to produce a bias of more than $50\%$ of the nominal value without triggering the BDD alarm. Biased estimates could directly lead to serious social and economical consequences. For instance, \cite{2012:Choi,2012:Jia,2011:Xie} showed that attackers equipped with data injection can manipulate the electricity price in power market. Worse still, \cite{2011:Yuan} warned that the attack can even cause regional blackout.

The conventional wisdom to mitigate false-data injection attack is by securing meter measurements to evade malicious injections, e.g. either by guards, video monitoring or tamper-proof communication system, etc \cite{2010:Sandberg,2010:Bobba,2012:Vukovic,2013:Giani,2012:Sou}. For instance, \cite{2010:Sandberg} quantifies the vulnerability of meter measurements in the presence of injection attack using ``security indices". \cite{2012:Sou} computes the critical points in the measurement set, the compromise of which would result in successful undetectable attack. \cite{2013:Giani} proposed to use phasor measurement units (PMUs), which provide direct voltage phasor angle measurements at the buses installed, to mitigate sparse undetectable attack. In particular, \cite{2010:Bobba} proved that it is necessary and sufficient to protect a set of \emph{basic measurements} so that no undetectable false-data injection attack can be launched, where the size of a set of basic measurements is the same as the number of unknown state variables in the state estimation problem. With a limited budget but the vast size of large-scale power networks, it is often not possible to completely eliminate the chance of undetectable attack. In practice, the system operator should first protect the state variables that have greater social/economic impact once compromised, such as those for critical buses/substations connected to heavily loaded or economically important areas, or with critical interconnection purposes \cite{2008:Chakrabarti}. Our preliminary study in \cite{2011:Bi} has partly addressed this issue, where a sequential method is proposed to find a minimum set of meter measurements for the protection of any set of state variables. However, the enumeration-based method is of very high computational complexity in large scale power systems. It is therefore meaningful to devise an efficient method to protect a subset of state variables that serves our best interests, and opens to the possibility of expanding the set of protected state variables in the future.

Another newly emerged approach against injection attack is to limit the attacker's knowledge of the topological information needed for performing attacks \cite{2012:Morrow,2012:Talebi}. In practice, this incurs lower operation cost as it merely needs to change the parameters of some software/hardware configurations, such as the adjustable transformer taps in transmission lines, or the flexible a.c. transmission systems (FACTS) to adjust the real-time effective reactance, etc \cite{2002:Nedic,2006:Meier}. It was shown in \cite{2012:Morrow} that the intentional topology perturbation enables the system operator to detect the presence of false-data injection using conventional residual test. However, a random topology perturbation does not fully eliminate the possibility of undetectable attack. For instance, \cite{2012:Rahman} showed that undetectable attack is still possible if the attacker has imperfect but structured topological information. Currently, it lacks of a systematic study that provides an explicit guideline for the system operator to efficiently utilize the \emph{covert topological information} against injection attacks.

In this paper, we focus on using covert topological information (CTI) to mitigate false-data injection attacks. By keeping the exact reactance of a minimum set of transmission lines from attackers, no undetectable injection attack can be formulated to compromise any set of state variables. Besides, we also propose a \emph{mixed defending strategy} that jointly considers CTI and conventional secure meter measurement methods. Our detailed contributions are listed as follows,
\begin{itemize}
  \item We derive from the attackers' perspective a necessary and sufficient condition to perform undetectable attack with partial topological information. In particular, we develop a \emph{min-cut} method to design the optimal attack, which requires the minimum knowledge of system topology. The result is useful to develop effective countermeasures against undetectable attack.
  \item We show that the solution to the optimal state variable protection, which defends a set of critical state variables with minimum CTI, can be obtained by solving a standard \emph{Steiner tree problem}. Although the problem is \textit{NP-hard}, many well-investigated exact and approximation algorithms can be directly applied.
  \item We also develop a mixed defending strategy that jointly considers the use of CTI and secure meter measurements when either method alone is costly or unable to achieve the protection objective. A mixed integer linear programming (MILP) formulation is introduced to obtain the optimal mixed defending strategy. To tackle the \emph{NP-hardness} of the problem, a tree pruning-based heuristic is further presented to produce an approximate solution in polynomial time.
\end{itemize}

\subsection{Related works}
State estimation protection is closely related to the concept of \emph{power network observability}. The conventional power network observability analysis studies whether a unique estimate of all unknown state variables can be determined from the measurements \cite{2004:Abur}. From the attacker's perspective, \cite{2010:Bobba} proved that an undetectable attack can be formulated if removing the measurements it compromises will make the power system unobservable. Conversely, \cite{2010:Kosut} showed that no undetectable attack can be formulated if the power system is observable from the protected meter measurements. The early work by Krumpholz \emph{et al}. \cite{1980:Krumpholz} stated that a power system is observable if and only if it contains a spanning tree, which we refer to as an edged-measured spanning tree, that satisfies certain measurement-to-transmission-line mapping rules. Few recent papers also applied graphical methods to study the attack/defending mechanisms of false-data injection. For instance, based on the results in \cite{1980:Krumpholz}, \cite{2011:Kosut} proposed an algorithm to quantify the minimum-effort undetectable attack, i.e. the non-trivial attack that compromises least number of meters without being detected. Besides, \cite{2011:Sou} used a min-cut relaxation method to calculate the security indices defined in \cite{2010:Sandberg} to quantify the resistance of meter measurements in the presence of injection attack. Similar min-cut approach was also applied in \cite{2012:Sou} to identify the critical points in the measurement set, the loss of which would render the power system unobservable.

In this paper, we also study the state estimation protection problem from a graphical perspective. The novelties of the proposed methods are in twofold. First, in addition to the conventional method of securing meter measurement, we introduce a \emph{new degree of freedom} of using CTI for system protection and derives explicit protection procedures. Second, we consider a more general problem of protecting any subset of state variables with minimum cost. The graphical method of using only CTI for state variable protection is proved to be equivalent to a well-studied Steiner tree problem. A more challenging problem is the mixed defending strategy that jointly considers the CTI and secure meter measurement protection. We formulate the problem into a variant Steiner arborescence problem and propose both exact and approximation graphical algorithms.

The rest of this paper is organized as follows. In Section II, we introduce some preliminaries about state estimation and false-data injection attack. We study the design of partial knowledge attack in Section III. In Section IV, we propose efficient defending mechanisms and discuss some application scenarios. The performance of the proposed defending mechanisms is evaluated in Section V. Finally, the paper is concluded in Section VI. An early version of this paper with only CTI protection was published in the $2013$ IEEE Globecom \cite{2013:Bi1}.

\section{Preliminary}
\subsection{DC measurement model and state estimation}
\begin{figure}
\centering
  \begin{center}
    \includegraphics[width=0.35\textwidth]{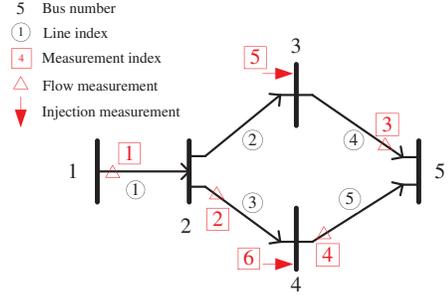}
  \end{center}
  \caption{Measurement placement of a 5-bus system with $v_5$ being the reference.}
  \label{61}
\end{figure}

We consider the linearized power network state estimation problem in a steady-state power system with $n+1$ buses and $t$ transmission lines. The topology of the power system can be characterized by a $t\times \left(n+1\right)$ incidence matrix $\mathbf{A}$ in a digraph. In this paper, we use the terms of buses and vertices, as well as transmission lines and edges, interchangeably. Let $\mathcal{V}$ and $\mathcal{E}$ denote the set of all vertices and edges, respectively. An entry $\left[\mathbf{A}\right]_{ij}=1$ indicates that edge $e_i$ leaves vertex $v_j$, where $v_j$ is referred to as the tail of $e_i$, denoted by $e^{(t)}_i$. $\left[\mathbf{A}\right]_{ij}=-1$ if $e_i$ enters vertex $v_j$, where $v_j$ is the head of $e_i$, denoted by $e^{(h)}_i$. The direction from $e^{(t)}_i$ to $e^{(h)}_i$ is the positive direction of $e_i$. $\left[\mathbf{A}\right]_{ij}=0$ if $e_i$ is not incident to $v_j$. Using a $5$-bus system in Fig. $\ref{61}$ for example,
\begin{equation}
\mathbf{A}=\left(
  \begin{array}{ccccc}
    1 & -1 & 0  & 0  & 0\\
    0 & 1 & -1 & 0  & 0\\
    0 & 1 & 0  & -1 & 0\\
    0 & 0 & 1  & 0  & -1\\
    0 & 0 & 0  & 1  & -1\\
  \end{array}
\right).
\end{equation}

The states of the power system include the bus voltage phase angles and voltage magnitudes. The voltage magnitudes can often be directly measured, while the values of phase angles need to be obtained from state estimation \cite{1994:Grainger}. \footnote{We also discuss including PMUs, which directly measures voltage phase angles, in the state estimation protection in Section IV.E.} In the linearized (DC) measurement model, we assume the knowledge of voltage magnitudes (i.e. $1$ in the per-unit system) at all buses and estimate the phase angles based on the active power measurements, i.e. the active power flows along the power lines and active power injections at buses \cite{2004:Abur}. Assume that a power system is measured by $m$ meters, including $m_F$ flow meters and $m_I$ injection meters. The set of all the meters are denoted by $\mathcal{M}$. Mathematically, the flow and injection measurements are related to phase angles as
\begin{equation}
\label{13}
\left(\begin{array}{c}
  \mathbf{F} \\
  \mathbf{I}
\end{array}\right) = \left(
               \begin{array}{c}
                  \mathbf{L}_F \mathbf{Y} \mathbf{A}\\
                  \mathbf{L}_I \mathbf{A}^\top \mathbf{Y} \mathbf{A}
               \end{array}
             \right) \boldsymbol{\theta_v} \triangleq \mathbf{H}\boldsymbol{\theta_v},
\end{equation}
where $\mathbf{H}$ is the measurement Jacobian matrix, $\boldsymbol{\theta_v}$ is the vector of all the phase angles. $\mathbf{Y}$ is a $t\times t$ diagonal matrix, where $\left[\mathbf{Y}\right]_{ii}\triangleq y_{e_i}$ is the reciprocal of the reactance of power line $e_i$. $\mathbf{L}_F$ is a $m_F\times t$ matrix, where $\left[\mathbf{L}_F\right]_{ij}=1$ if the $i^{th}$ flow meter measures the power flow in the positive direction of $e_j$ ($F^+_{e_j}$), $-1$ if it measures the negative direction ($F^-_{e_j}$) and $0$ otherwise. $\mathbf{L}_I$ is a $m_I\times (n+1)$ matrix with $\left[\mathbf{L}_I\right]_{ij}=1$ indicating that the $i^{th}$ injection meter measures the power injection at the $j^{th}$ bus ($I_{v_j}$). The superscript $(\cdot)^\top$ denotes the transpose operation. Using the power system in Fig. $\ref{61}$ for example,
\begin{equation}
\mathbf{L}_F=\left(
  \begin{array}{ccccc}
    1 & 0 & 0  & 0  & 0\\
    0 & 0 & 1  & 0  & 0\\
    0 & 0 & 0  & -1  & 0\\
    0 & 0 & 0  & 0  & 1\\
  \end{array}
\right),
\mathbf{L}_I=\left(
  \begin{array}{ccccc}
    0 & 0 & 1  & 0  & 0\\
    0 & 0 & 0  & 1  & 0\\
  \end{array}
\right).
\end{equation}
Specifically, the rows of $\mathbf{L}_F$ correspond to flow meters $\left\{r_1,r_2,r_3,r_4\right\}$ and the columns correspond to lines $e_1$ to $e_5$. The rows of $\mathbf{L}_I$ correspond to injection meters $\left\{r_5,r_6\right\}$, and the rows correspond to buses $v_1$ to $v_5$. Suppose that the reactance of all transmission lines equals $1$,
\begin{equation}
\mathbf{H}=\left(
  \begin{array}{ccccc}
    1 & -1 & 0  & 0  & 0\\
    0 & 1 & 0 & -1  & 0\\
    0 & 0 & -1  & 0  & 1\\
    0 & 0 &  0  & 1  & -1\\
    0 & -1 & 2  & 0 & -1\\
    0 & -1 & 0  &  2 & -1\\
  \end{array}
\right),
\end{equation}
where the first $4$ rows represent flow measurements while the last two rows represent injection measurements.

Based on the above measurement model, we introduce the following definition $1$ on measurability.

\textbf{Definition $1$: (measurability)} A flow meter on transmission line $e_i$ (corresponding to $F^+_{e_i}$ or $F^-_{e_i}$), \emph{measures} the edge $e_i$, and the two vertices of $e_i$, i.e. $e^{(h)}_i$ and $e^{(t)}_i$. An injection meter on bus $v_j$ \emph{measures} the edges connected to the buses, i.e. $\left\{e_i\mid e_i\in \mathcal{N}^+_j \cup \mathcal{N}^-_j\right\}$, and the vertices of the measured edges, i.e. $\left\{e^{(h)}_i,e^{(t)}_i \mid e_i\in \mathcal{N}^+_j \cup \mathcal{N}^-_j\right\}$, where $\mathcal{N}^+_j$ ($\mathcal{N}^-_j$) is the set of edges that $v_j$ being the tail (head).

For instance, the flow meter $r_4$ measures edge $e_5$ and vertices $v_4$ and $v_5$. The injection meter $r_5$ measures edges $e_2$ and $e_4$ that connects to $v_3$, and vertices $v_2$, $v_3$ and $v_5$. There could be the case that some components are not measured by either flow or injection meters, such as edge $e_4$ in Fig. \ref{71}. As we will show in the following section, the state estimation protection problem is only related to the measured vertices and edges, thus all the unmeasured elements can be discarded without affecting the problem solution.

Without loss of generality, we assume that bus $n+1$ is the reference bus, denoted by $R$, whose phase angle equals zero. The state estimation problem is therefore to estimate the other $n$ phase angle state variables, which is captured by the vector $\boldsymbol{\theta}=\left(\theta_1,\theta_2,..,\theta_n\right)^\top$, based on the $m$ measurements $\mathbf{z}=\left(z_1,z_2,..,z_m\right)^\top$, where
\begin{equation}
\label{18}
\mathbf{z} = \mathbf{\bar{H}}\boldsymbol{\theta} + \mathbf{e}.
\end{equation}
Here $\mathbf{\bar{H}}$ is the reduced measurement Jacobian matrix excluding the column that corresponds to the reference bus in $\mathbf{H}$. $\mathbf{e}\thicksim \mathcal{N}\left(\mathbf{0},\mathbf{Q}\right)$ is independent measurement noise, where $\mathbf{Q}$ is the diagonal covariance matrix. When $\mathbf{\bar{H}}$ is full column rank, i.e. $rank\left(\mathbf{\bar{H}}\right)=n$, the maximum likelihood estimate $\boldsymbol{\hat{\theta}}$ is
\begin{equation}
\label{3}
\boldsymbol{\hat{\theta}} = \left(\mathbf{\bar{H}}^{T}\mathbf{Q}^{-1}\mathbf{\bar{H}}\right)^{-1}\mathbf{\bar{H}}^{T}\mathbf{Q}^{-1}\mathbf{z} \triangleq \mathbf{Pz}.
\end{equation}
In this case, all the state variables have an unique estimate and the power system is called observable from the measurement set $\mathcal{M}$. In this paper, we assume that all the power systems in consideration are observable from the available measurements. Equivalently, the observability of a power system can be characterized in a graph \cite{1980:Krumpholz}.

\textbf{Proposition $1$:} The power network $G=\left(\mathcal{V},\mathcal{E}\right)$ is observable from its measurement set $\mathcal{M}$ if and only if the graph defined on $G$ contains a spanning tree, where each edge of which is mapped to a meter according to the following rules,
\begin{enumerate}
  \item an edge is mapped to a flow meter placed on it, if any;
  \item an edge without a flow meter is mapped to an injection meter that measures it;
  \item different edges are mapped to different meters in $\mathcal{M}$.
\end{enumerate}

\emph{Proof:} See \cite{1980:Krumpholz} for detailed proof.  $\hfill \blacksquare$

A spanning tree that satisfies the conditions in Proposition $1$ is referred to as an \emph{edge-measured spanning tree} (EMST). An observable power network may contain multiple EMSTs, which can be easily found using a max-flow method in polynomial time \cite{1986:Barglela}. Using Fig. \ref{61} for example, we can find two EMSTs with the following edge-to-measurement mappings:
\begin{enumerate}
  \item $\left\{e_1,e_2,e_4,e_5\right\}\leftrightarrow \left\{r_1,r_5,r_3,r_4\right\}$;
  \item $\left\{e_1,e_2,e_3,e_4\right\}\leftrightarrow \left\{r_1,r_5,r_6,r_3\right\}$.
\end{enumerate}
Essentially, each EMST corresponds to a set of basic measurements, from which the power network is observable. The properties of EMST are important to derive attacking and defensive strategies in later sections.

In an observable power network, the estimates $\boldsymbol{\hat{\theta}}$ could be wrong due to random measurement errors or attacks. Some previous studies detect erroneous estimation by assuming the prior knowledge of the distribution of state variables, e.g. the GLRT-based (generalized likelihood ratio test) detector proposed in \cite{2011:Kosut}. In this paper, we do not assume such knowledge and adopt the conventional BDD method, which compares the squared $l_2$-norm of measurement residual with a threshold $\tau$. The BDD identifies bad data measurements if
\begin{equation}
\label{4}
r=||\mathbf{z}-\mathbf{\bar{H}\boldsymbol{\hat{\theta}}}||= ||\left(\mathbf{I - \mathbf{\bar{H}P}}\right) \mathbf{e}||> \tau.
\end{equation}
Otherwise, $\mathbf{z}$ is considered as a normal measurement.

\subsection{Undetectable attacks and protection model}
Suppose that attackers inject data $\mathbf{a}=\left(a_1,a_2,..,a_m\right)^\top$ into measurements. Then, the received measurements become
\begin{equation}
\label{5}
\mathbf{\tilde{z}} = \mathbf{\bar{H}}\boldsymbol{\theta} + \mathbf{e} +\mathbf{a}.
\end{equation}
In general, an unstructured $\mathbf{a}$ is likely to be identified by the BDD. However, it is found in \cite{2009:Liu} that some well-structured data injections, such as those with $\mathbf{a}=\mathbf{\bar{H}c}$, can bypass BDD. Here $\mathbf{c} = \left(c_1,c_2,..,c_n\right)^\top$ is a random vector. This can be verified by calculating the measurement residual in (\ref{5}), where
\begin{equation}
\label{11}
\tilde{r} = ||\mathbf{\tilde{z}}- \mathbf{\bar{H}P\tilde{z}} || = ||\mathbf{z}+\mathbf{a}-\mathbf{\bar{H}}(\mathbf{\boldsymbol{\hat{\theta}}+c})||= ||\mathbf{z}-\mathbf{\bar{H}\boldsymbol{\hat{\theta}}}||.
\end{equation}
Same residual is obtained as if no malicious data were injected. Therefore, a structured attack $\mathbf{a}=\mathbf{\bar{H}c}$ will not be detected by BDD. We refer to such an injection attack with the residual in (\ref{11}) as an \emph{undetectable attack}. In this case, the system operator would mistake $\boldsymbol{\hat{\theta}}+\mathbf{c}$ for a valid estimate, and thus an error vector $\mathbf{c}$ is introduced.

We see that attackers require decent knowledge of $\mathbf{\bar{H}}$ to perform an undetectable attack. Conversely, the system operator has the potential to eliminate the chance of undetectable attacks by limiting attackers' knowledge. From (\ref{13}), the reduced measurement Jacobian matrix is
\begin{equation}
\mathbf{\bar{H}} = \left(
               \begin{array}{c}
                  \mathbf{L}_F \mathbf{Y} \mathbf{\bar{A}}\\
                  \mathbf{L}_I \mathbf{A}^\top \mathbf{Y} \mathbf{\bar{A}}
               \end{array}
             \right),
\end{equation}
where $\mathbf{\bar{A}}$ is the submatrix of $\mathbf{A}$ excluding the column of the reference bus. From a system operator's perspective, we assume a well-informed attacker with the perfect knowledge of $\mathbf{L}_F$, $\mathbf{L}_I$ and $\mathbf{\bar{A}}$. However, it has imperfect knowledge of $\mathbf{Y}$. One reason is that the system operator can secretly vary $\mathbf{Y}$ by adjusting the transformer taps installed on the transmission lines, or using the FACTS to adjust the real-time effective reactance \cite{2002:Nedic,2006:Meier}. We assume the system operator can keep from attackers the exact reactance of a set of transmission lines $\mathcal{K}\subseteq \mathcal{K}_0$, where $\mathcal{K}_0$ is the set of lines that the system operator has the capability to protect. Its objective is to ensure that no undetectable attack can be formulated to compromise a given set of state variables $\mathcal{D}\subseteq \mathcal{I}$, where $\mathcal{I}$ is the set of all unknown state variables. That is, $c_i=0$ for all $i\in \mathcal{D}$. In some cases, protecting topological information alone may fail to achieve the protection objective, e.g. the situation introduced in Section III.A. In general, a mixed defending strategy is needed to combine the CTI with secure measurement method. Then, the question is how to defend $\mathcal{D}$ with minimum cost on line information protection and meter measurement security.

\section{Optimal Undetectable Attack with Partial Topological Knowledge}
To understand the vulnerability of the power system, we investigate in this section from the attackers' perspective the design of an undetectable attack when full knowledge of topological information ($\mathbf{Y}$) is not available. In Section III.A, we first show that undetectable attack can even be performed without any knowledge of $\mathbf{Y}$ when the measurement placement has a special structure, i.e. contains a bridging edge. In a system free of bridging edges, we show in Section III.B that undetectable attack is still possible if the attacker has a structured partial knowledge of $\mathbf{Y}$. Then, we formulate the optimal attack, which compromises a given set of state variables with minimum knowledge of network topology, into a min-cut problem. These results will be used in the next section to derive defending mechanisms using covert topological information.

\subsection{Undetectable attack in a system containing bridging edges}
For the moment, we assume that no meter is secured by the system operator. The imperfect knowledge of $\mathbf{Y}$ at the attacker is denoted by $\mathbf{\tilde{Y}}=\mathbf{Y}+ \boldsymbol{\epsilon}$, where $\boldsymbol{\epsilon}$ is a diagonal error matrix unknown to the attackers caused by system operator's countermeasures, and $\epsilon_{e_l}=0$ if the attacker has the perfect knowledge of line $e_l$. Then, the attacker's knowledge of $\mathbf{\bar{H}}$ is
\begin{equation}
\label{12}
\mathbf{\tilde{H}} \triangleq \mathbf{\bar{H}} + \boldsymbol{\delta} =\mathbf{\bar{H}}  + \left(
               \begin{array}{c}
                  \mathbf{L}_F \boldsymbol{\epsilon} \mathbf{\bar{A}}\\
                  \mathbf{L}_I \mathbf{A}^\top \boldsymbol{\epsilon} \mathbf{\bar{A}}
               \end{array}
             \right).
\end{equation}
If the attacker constructs an injection based on the biased measurement Jacobian as $\mathbf{a} = \mathbf{\tilde{H}} \mathbf{c}$, the residual norm becomes
\begin{equation}
\label{31}
\tilde{r}= ||\mathbf{\tilde{z}}- \mathbf{\bar{H}P\tilde{z}}|| = ||\left(\mathbf{I - \mathbf{\bar{H}P}}\right)  \boldsymbol{\delta} \mathbf{c} + \left(\mathbf{I - \mathbf{\bar{H}P}}\right) \mathbf{e}||.
\end{equation}
The residual due to attack is $\left(\mathbf{I - \mathbf{\bar{H}P}}\right)  \boldsymbol{\delta} \mathbf{c}$. Meanwhile, the estimate of $\boldsymbol{\theta}$ is
\begin{equation}
\label{32}
\boldsymbol{\tilde{\theta}} = \mathbf{P\tilde{z}} = \hat{\boldsymbol{\theta}} + \mathbf{c} + \mathbf{P} \boldsymbol{\delta} \mathbf{c}.
\end{equation}

To perform an undetectable attack, we see that $\boldsymbol{\delta} \mathbf{c}$, or equivalently the attack vector $\mathbf{a}$, must lie in the null space of $ \mathbf{R} \triangleq \mathbf{I - \mathbf{\bar{H}P}}$. In general, the attacker does not know the exact $\mathbf{Y}$ (and thus exact $\mathbf{\bar{H}}$ and $\mathbf{P}$), therefore cannot obtain a nontrivial null space of $\mathbf{R}$. However, we show in the following that simple null space solutions of $\mathbf{R}$ can be found without knowing $\mathbf{Y}$ if the measurement placement contains a bridging edge structure defined as follows.

\textbf{Definition $2$: (bridging edge)} In the graph $G=\left(\mathcal{V},\mathcal{E}\right)$ defined on a power network, an edge is a \emph{bridging edge} if it is contained in all edge-measured spanning trees (EMSTs) obtained from the power system measurement placement.

\begin{figure}
\centering
  \begin{center}
    \includegraphics[width=0.4\textwidth]{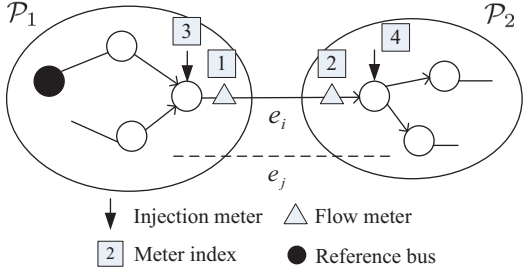}
  \end{center}
  \caption{An illustration of a bridging edge $e_i$.}
  \label{62}
\end{figure}

For instance, $e_1$ in Fig. \ref{61} is an bridging edge since all EMSTs must contain $e_1$ to connect $v_1$. In practice, a well-connected power network with many redundant meter measurements contain only few bridging edges. For example, the measurement placement of IEEE $14$-bus system in Fig. \ref{71} has only one bridging edge $e_{14}$. The existence of a bridging edge indicates poor connectivity from some vertices to the reference vertex in the EMSTs. To see this, we consider a bridging edge $e_i$ in an EMST in Fig. \ref{62}. The two partitions of vertices connected by $e_i$ are denoted by $\mathcal{P}_1$ and $\mathcal{P}_2$. Without loss of generality, we assume that the reference vertex is in $\mathcal{P}_1$. In fact, we can infer that all the EMSTs observe the same vertex partitions $\mathcal{P}_1$ and $\mathcal{P}_2$. Otherwise, another $e_j$ connecting the two partitions in an EMST would result in a circle. In particular, all the vertices in $\mathcal{P}_2$ will be separated from the reference vertex if $e_i$ is removed in the EMSTs. After taking all the bridging edges into consideration, we can categorize the vertices into two non-overlapping types, determined by the connectivity to the reference vertex as stated in Definition $3$.

\textbf{Definition $3$: (vertex type)} A vertex is referred to as a $\mathcal{P}_2$-type vertex if it is separated from the reference vertex in any EMST after all the bridging edges are removed. Otherwise, it is a $\mathcal{P}_1$-type vertex.

For instance, $v_1$ in Fig. \ref{61} and $v_8$ in Fig. \ref{71} are $\mathcal{P}_2$-type vertices, while all the other vertices are $\mathcal{P}_1$-type in the respective cases. Next, we establish in the following Proposition $2$ the connection between bridging edge and undetectable attack without the knowledge of topological information.

\textbf{Proposition $2$:} If a measurement placement contains bridging edge(s), an undetectable attack can be performed to compromise $\mathcal{P}_2$-type vertices without the knowledge of $\mathbf{Y}$.

\emph{Sketch of proof:} Consider the bridging edge $e_i$ in Fig. \ref{62}, which is mapped to either a flow meter ($r_1$ or $r_2$), or an injection meter at either end of the edge ($r_3$ or $r_4$), following the definition of an EMST. Without loss of generality, we assume that all the four meters are available in the measurement set. Then, it can be easily verified that a simple attack vector $\mathbf{a} = \left[1,-1,1,-1\right]^\top$ to $r_1$ to $r_4$, which injects no data to other meters, can cause a decrease of $1/y_{e_i}$ to all the state variables in $\mathcal{P}_2$ and no impact to the state variables in $\mathcal{P}_1$, and the residual due to attack is zero. This completes the proof. $\hfill \blacksquare$

\emph{Remark $1$:} Although the attack in Proposition $2$ is undetectable, the attacker does not know the exact magnitude of bias it causes as $y_{e_i}$ is unknown.

Using Fig. \ref{61} for example, a simple undetectable attack that increases the reading of meter $r_1$ by $1$ unit can compromise the  $\mathcal{P}_2$-type vertex $v_1$ without knowing $\mathbf{Y}$. Essentially, this is because $\mathbf{R}$ has special structures to be explored when a bridging edge exists. Let $y_{e_i}$ in Fig. \ref{61} be randomly generated for each line. An example realization of the symmetric matrix $\mathbf{R}$ is
\begin{equation*}
\mathbf{R}=\left(
  \begin{array}{cccccc}
    0 & 0 & 0  & 0   & 0 & 0\\
    0 & 0.468  & 0.284  & -0.195  & 0.128 & 0.337\\
    0 & 0.284  & 0.602  & 0.292   & 0.271 & 0.007\\
    0 & -0.195 & 0.292  & 0.475   & 0.131 & -0.330\\
    0 & 0.128  & 0.271  & 0.131   & 0.122 & 0.003\\
    0 & 0.337  & 0.007  & -0.330  & 0.003 & 0.334\\
  \end{array}
\right).
\end{equation*}
We notice that all the entries in the first row (column) of $\mathbf{R}$ are always zero regardless of the value of $\mathbf{Y}$. Therefore, $\mathbf{a}=[1,0,0,0,0,0]^\top$ satisfies $\mathbf{Ra}=\mathbf{0}$, i.e. zero residual.

In practice, bridging edges, and thus $\mathcal{P}_2$-type vertices, can be easily identified by attackers either by checking the network topology and measurement placement, or directly observing the null space of $\mathbf{R}$ as $\mathbf{Y}$ randomly varies. Then, following Proposition $2$, undetectable attacks can be performed to compromise $\mathcal{P}_2$-type vertices without knowing $\mathbf{Y}$. However, constructing an undetectable attack to compromise the remaining $\mathcal{P}_1$-type vertices is much more difficult. Equivalently, the attack vector can be obtained by constructing an undetectable attack in a residual power system by removing all the $\mathcal{P}_2$-type vertices. In such a power system free of bridging edges, the knowledge of $\mathbf{Y}$ is critical to construct undetectable attack vectors, as the null space of $\mathbf{R}$ for the residual power system depends on the value of $\mathbf{Y}$. In this case, the attacking strategy in Fig. \ref{62} to a non-bridging edge $e_j$ becomes detectable. Intuitively, this is because the state estimates derived from two different EMSTs can be inconsistent in the presence of attack, if one tree uses the edge $e_j$ mapped to a compromised meter while the other does not use $e_j$ at all.

To study the method to compromise $\mathcal{P}_1$-type vertices, we focus on systems without bridging edges in the remaining part of this section. In particular, we show that undetectable attacks is still possible when the attacker has limited but structured partial knowledge of $\mathbf{Y}$. With a bit abuse of notations, we still use the previously defined variables and parameters, such as $\mathbf{A}$, $\mathbf{\bar{H}}$ and $\mathbf{c}$, to denote the corresponding items in the residual power system free of bridging edges, unless stated otherwise.

\subsection{Undetectable attack in a system without bridging edges}
In a power system free of bridging edges, we first show that a necessary and sufficient condition to launch an undetectable attack is $\boldsymbol{\delta} \mathbf{c}=\mathbf{0}$ and $\mathbf{c}\neq \mathbf{0}$. For the sufficient argument, we see from (\ref{32}) that a nonzero error $\mathbf{c}$ is introduced to $\mathbf{\hat{x}}$ if $\boldsymbol{\delta} \mathbf{c}=\mathbf{0}$. Besides, the residual in (\ref{31}) becomes the same as in (\ref{4}), as if no malicious data is injected. The necessary argument is because a nontrivial null space of $\mathbf{R}$ cannot be obtained by attackers without knowing the exact $\mathbf{\bar{H}}$ or $\mathbf{P}$. From (\ref{12}), the attacker must ensure the following conditions
\begin{equation}
\label{33}
\begin{aligned}
\mathbf{L}_F \boldsymbol{\epsilon} \mathbf{\bar{A}c} = \mathbf{0},\ \ \  \mathbf{L}_I \mathbf{A}^\top \boldsymbol{\epsilon} \mathbf{\bar{A}c} =\mathbf{0}, \ \ \ \mathbf{c}\neq \mathbf{0}.
\end{aligned}
\end{equation}
Equivalently, we have
\begin{subequations}
\label{34}
\begin{align}
&\left[\mathbf{L}_F\right]_{il}\epsilon_{e_l} \left[\mathbf{\bar{A}}\right]_{l*}\mathbf{c}= 0,\ \forall i, l\label{35}\\
&\left[\mathbf{L}_I\right]_{ij} \sum_{l=1}^L \left[\mathbf{A}\right]_{lj} \epsilon_{e_l} \left[\mathbf{\bar{A}}\right]_{l*}\mathbf{c}=0, \ \forall i, j,\label{36}
\end{align}
\end{subequations}
where $\left[\mathbf{\bar{A}}\right]_{l*}$ denotes the $l^{th}$ row of matrix $\mathbf{\bar{A}}$.

\begin{figure}
\centering
  \begin{center}
    \includegraphics[width=0.4\textwidth]{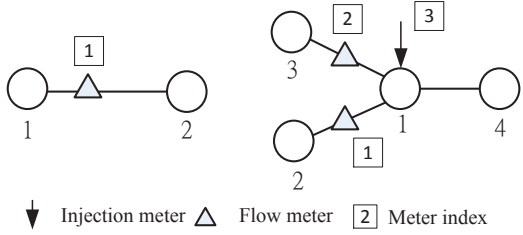}
  \end{center}
  \caption{Illustration of undetectable attack with partial network information.}
  \label{64}
\end{figure}

We illustrate in Fig. $\ref{64}$ a simple case that an undetectable attack satisfies (\ref{34}). From (\ref{35}), an undetectable attack to compromise a flow measurement must satisfy
\begin{equation}
\label{40}
\epsilon_{12}\left(c_1-c_2\right)=0.
\end{equation}
From (\ref{36}), an attack to compromise an injection measurement placed at bus $1$ must satisfy
\begin{equation}
\epsilon_{12}\left(c_1-c_2\right) + \epsilon_{13}\left(c_1-c_3\right) +  \epsilon_{14}\left(c_1-c_4\right)=0.
\end{equation}
Since $\epsilon_{1i}$'s are unknown random errors, the attacker must force each individual term to be $0$, i.e.
\begin{equation}
\label{41}
\epsilon_{1i}\left(c_1-c_i\right)=0,\ i=2,3,4.
\end{equation}
In other words, (\ref{36}) can be decomposed into a number of flow measurement conditions defined on the transmission lines measured by the injection meter $i$, i.e.
\begin{equation}
\left[\mathbf{L}_I\right]_{ij} \left[\mathbf{A}\right]_{lj} \epsilon_{e_l} \left[\mathbf{\bar{A}}\right]_{l*}\mathbf{c}=0, \ \forall i, j, l.
\end{equation}
From (\ref{40}) and (\ref{41}), a measured edge $[i,j]$ must satisfy either $\epsilon_{ij}=0$ or $c_i=c_j$, or both. Formally, we specify in the following Theorem $1$ the necessary and sufficient condition to perform an undetectable attack.

\textbf{Theorem $1$:} For a power system measurement placement free of bridging edges, an undetectable attack can be performed if and only if each measured transmission line $e_l$, either by a flow measurement ($\left[\mathbf{L}_F\right]_{il} \neq 0$ for some $i$) or injection measurement ($\left[\mathbf{L}_I\right]_{ij} \left[\mathbf{A}\right]_{lj}\neq 0$ for some $i$ and $j$), satisfies at least one of the following two conditions
\begin{enumerate}
  \item $\epsilon_{e_l}=0$, i.e. perfect knowledge of transmission line $e_l$,
  \item $c_{e_l^{(h)}}=c_{e_l^{(t)}}=\beta$. That is, the same error $\beta$ is introduced to the head and tail vertices of $e_l$, where $\beta$ is an arbitrary real number.
\end{enumerate}

We first assume that the attacker intends to compromise a single state variable $\theta_k$. That is, $c_k=1$ and $c_i=0$, $\forall i\neq k$. The transmission lines that are not incident to bus $v_k$ automatically satisfy the condition $2$), i.e. $\beta=0$. Therefore, the attacker only needs to obtain the perfect knowledge of the measured transmission lines that are incident to $v_k$. Now, let us relax the attacker's objective to compromise $\theta_k$ regardless of its influence to the other state variables, i.e. $c_k=1$ only. In this case, besides obtaining the knowledge of a transmission line incident to bus $v_k$, the attacker can also satisfy condition $2$) by letting $c_i=1$ for the bus $v_i$ connected to $v_k$. In fact, the attacker can further introduce errors to the two-hop neighboring buses until an optimal solution, which requires the least knowledge of $\mathbf{Y}$, is obtained.

Conceptually, the attacker needs to separate the network into two disjoint subnetworks, where the same error $\beta=1$ is introduced to the buses in the subnetwork that includes the tagged bus $v_k$, and $\beta=0$ for the buses in the other subnetwork. Then, the attacker only needs to obtain the perfect knowledge of the measured transmission lines that connect the two subnetworks. It is worth noticing that the tagged bus $v_k$ and the reference bus $R$ cannot be included in the same subnetwork. Otherwise, the undetectable attack can not introduce any error to the tagged bus $v_k$ because the value of the reference bus is set to be $0$ by default. For instance, after removing bridging edge $e_1$ and bus $v_1$ in Fig. $\ref{61}$, a cut on edges $e_3$ and $e_4$ separates the residual power network into two disjoint subnetworks $\left\{v_2,v_3\right\}$ and $\left\{v_4,v_5\right\}$. Suppose that the attacker intends to compromise bus $v_3$, which is achievable by letting $c_2=c_3=1$ and $c_4=0$. After obtaining the perfect knowledge of $e_3$ and $e_4$, the attacker needs to inject to $\left\{r_2,r_3,r_4,r_5,r_6\right\}$ by $\mathbf{a}=\mathbf{\tilde{H}c} =\left[1,1,0,1,-1\right]$, where $\mathbf{c} = \left[c_2,c_3,c_4\right]$ is the state vector and $\mathbf{\tilde{H}}$ denotes the reduced measurement Jacobian known by the attacker after removing $e_1$. As bus $v_5$ is the reference bus, the state estimator in (\ref{3}) yields that the error introduced to bus $v_3$ is $1$. However, if $v_2$ is the reference bus, such that $v_3$ and the reference are now included in the same subnetwork, the error introduced to $v_3$ is $0$. By definition, the attack to compromise $v_3$ is failed.

In general, the attacker needs to separate the network into a number of disjoint subnetworks when it intends to compromise a set of state variables $\mathcal{D}$. Without causing confusions, we use $\mathcal{D}$ to denote the buses of the corresponding state variables. We denote the vertex sets in the subnetworks by $\mathcal{S}_0,..,\mathcal{S}_K$, where $R\in \mathcal{S}_0$ and $\mathcal{D}\subseteq \mathcal{S}\setminus \mathcal{S}_0$ with $\mathcal{S}=\mathcal{S}_0\cup\mathcal{S}_1\cup..\cup\mathcal{S}_K$ being the set of all vertices. Without loss of generality, an undetectable attack can be formulated by introducing error $\beta_i=i$ to state variables within $\mathcal{S}_i$ and obtaining the perfect knowledge of transmission lines that connect different subnetworks.

\begin{figure}
\centering
  \begin{center}
    \includegraphics[width=0.30\textwidth]{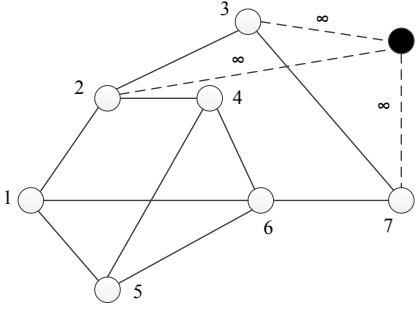}
  \end{center}
  \caption{Min s-t cut problem with $|\mathcal{D}|>1$. Here, $v_1$ is the reference bus, $\mathcal{D}=\left\{v_2,v_3,v_7\right\}$ and the shaded vertex denotes the supersink.}
  \label{53}
\end{figure}

\subsection{Optimal attack using min-cut method}
An optimal partial knowledge attack, which requires the minimum cost in acquiring necessary topological information, can be formulated by solving an s-t min-cut problem.

\textbf{Definition $4$: (Minimum s-t cut problem)} An s-t cut $C=(\mathcal{S},\mathcal{T})$ in an undirected graph $G=(\mathcal{V},\mathcal{E})$ is a partition of $\mathcal{V}$ such that $s\in \mathcal{S}$ and $t\in \mathcal{T}$, $\mathcal{S}\cup \mathcal{T}=\mathcal{V}$. The weight of the cut is the sum of the positive weights of edges between vertices in each part, where
\begin{equation}
w(\mathcal{S},\mathcal{T})=\sum_{u\in \mathcal{S},v\in \mathcal{T}} w_{uv}.
\end{equation}
The minimum s-t cut problem is to determine the cut $(\mathcal{S},\mathcal{T})$ such that the weight is minimized.

The min s-t cut problem seeks for the optimal cut that separates the source vertex $s$ and sink vertex $t$ with minimum edge cost. It can be efficiently solved using max-flow approach in polynomial time. The fastest maximum flow algorithm to solve the min s-t cut problem currently takes $O\left(|\mathcal{E}||\mathcal{V}|\log\left(|\mathcal{V}|^2/|\mathcal{E}|\right)\right)$ time complexity \cite{2012:Sou}, where $|\mathcal{V}|$ and $|\mathcal{E}|$ are the number of vertices and edges in the graph.

The optimal attack formulation can be easily converted into an s-t min cut problem when $|\mathcal{D}|=1$, where the reference bus is set to be the source $s$ and the only target bus is the sink $t$. The edge weight $w_{uv}$ represents the difficulty, measured in dollars, of obtaining the perfect knowledge of a measured line $\left[u,v\right]$, incorporating the factors such as geographical locations and level of protections, etc. In particular, $w_{uv}=\infty$ for those transmission lines whose line reactance is impossible to obtain, and $0$ for publicly accessible information. A slight modification is needed to solve the problem for $|\mathcal{D}|>1$. As illustrated in Fig. $\ref{53}$, this is achieved by adding a \emph{supersink} $t$ and connecting all the buses in $\mathcal{D}$ to $t$ through edges with infinite cost. Then, the optimal solution can be obtained by solving a standard min s-t cut problem that separates the reference and the supersink. The detailed procedures are summarized in Algorithm $1$. The attacker needs (and only needs) to inject false data to boundary meters, either flow or injection, that measure buses at both sides of $\mathcal{S}$ and $\mathcal{T}$.

In fact, our results are consistent with \cite{2011:Kosut} that an undetectable attack can be formulated by attacking the boundary meters in the cut that separates the power network. Here, instead of assuming the perfect knowledge of $\mathbf{H}$, we further show that an attack can be performed successfully with limited topological information. More importantly, the necessary condition to perform undetectable attack derived in Theorem $1$ can be used to devise effective countermeasures against undetectable attacks, which is detailed in the next section.

\begin{algorithm}
\small
 \SetAlgoLined
 \SetKwData{Left}{left}\SetKwData{This}{this}\SetKwData{Up}{up}
 \SetKwRepeat{doWhile}{do}{while}
 \SetKwFunction{Union}{Union}\SetKwFunction{FindCompress}{FindCompress}
 \SetKwInOut{Input}{input}\SetKwInOut{Output}{output}
 \Input{$\mathcal{I}, \mathcal{D}$, $R$, edge weight vector $\mathbf{w}$.}
 \Output{attacking vector $\mathbf{a}$ to compromise $\mathcal{D}$}
      Construct a weighted undirected graph $G=(\mathcal{V},\mathcal{E})$ by removing unmeasured edges and assigning weights to all the remaining edges\;
      Choose the reference bus as the source $s$. Add a supersink $t$, which connects to all the buses in $\mathcal{D}$ with infinite edge weight. Find the minimum s-t cut, denoted by $C=(\mathcal{S},\mathcal{T})$\;
      Obtain the exact reactance of the transmission lines in the cut. That is, $\epsilon_{ij}=0$ for $i\in \mathcal{S}$ and $j\in \mathcal{T}$. Introduce the error $\beta$ to all the state variables in $\mathcal{T}$, i.e. $c_j=\beta$ for all $j\in \mathcal{T}$. Besides, $c_i=0$ for all $i\in\mathcal{S}$\;
      Inject attacking vector $\mathbf{a}=\mathbf{\tilde{H}c}$ to boundary measurements.
 \caption{Minimum-cost partial knowledge attack}
\end{algorithm}

\section{State Variable Protection via Covert Topological Information}
In this section, we propose the methods to defend any given set of state variables against undetectable attacks. Specifically, we characterize the optimal CTI protection problem as a well-studied Steiner tree problem. We further propose a mixed defending strategy that jointly considers the CTI protection and conventional meter measurement protection methods. Both exact and approximate algorithms are proposed to obtain a solution of the mixed defending strategy.

\subsection{Pure CTI protection}
We have shown in Section III.A that undetectable attack can be constructed to compromise $\mathcal{P}_2$-type vertices without knowing $\mathbf{Y}$. It is therefore hopeless to protect $\mathcal{P}_2$-type vertices by keeping transmission line information covert. Alternatively, $\mathcal{P}_2$-type vertices can be protected by conventional secure meter measurement method. For the moment, we assume the power system is free of bridging edges, and exploit the solution structure of using pure CTI to defend against undetectable attacks. Based on the result, a mixed defending strategy using both CTI and secure meter measurements is developed in the next subsection to incorporate the presence of bridging edges.

From Section III.C, to compromise a set of state variables $\mathcal{D}$, attackers need to obtain the perfect knowledge of a set of transmission lines, which eventually forms a cut that separates the buses correspond to $\mathcal{D}$ and the reference bus. Conversely, to prevent any undetectable attack from compromising $\mathcal{D}$, the system operator needs to maintain ``covert" paths linking the reference bus to all the buses correspond to $\mathcal{D}$. This is formally proved in the following Theorem $2$.

\textbf{Theorem $2$: } In a power system measurement placement free of bridging edges, no undetectable attack can be formulated to compromise a set of state variables $\mathcal{D}$ if and only if the graph $G$ contains a tree that connects the reference bus with all the vertices in $\mathcal{D}$. Each edge of the tree is measured and its line admittance is covert from the attackers.

\emph{Proof:} We first show the $\emph{if}$ part. For each bus $v_k\in\mathcal{D}$, there exists a covert path from $R$ to $v_k$, consisting of covert lines. Without loss of generality, we denote the indices of buses in the path, from $R$ to the tagged bus $v_k$, by $\left\{0,1,2,\cdots,p\right\}$. Since $c_0=0$ (the default value of $R$), we have $c_1=0$. This is because introducing non-zero error to bus $1$ will change the readings of power flow in edge $\left[0,1\right]$ (the reading could be of an injection meter), which is not achievable under the assumption that $y_{01}$ is not perfectly known to the attacker and $e_{[0,1]}$ is not a bridging edge. Then, we can argue inductively that $c_i=c_{i-1}$ for $i=2,\cdots,p$. Therefore, we have $c_1=\cdots =c_k=0$. In other words, no error can be introduced to any bus $v_k\in\mathcal{D}$.

Then, we prove the \emph{only if} part. If a vertex in $\mathcal{D}$ is not included in the tree, attacker can always find a cut that separates the bus and the reference bus. Thus, an undetectable attack can be formulated following the steps given in Algorithm $1$. $\hfill \blacksquare$

Theorem $2$ indicates that the optimal defending strategy is equivalent to finding a minimum Steiner tree (MST) which connects all vertices in $\mathcal{D}\cup R$ with minimum edge weight sum. The weight of an edge, measured in dollars, is the cost of protecting the corresponding transmission line information. An illustrative example is provided in Section V.B and we do not present here for brevity. A number of exact and approximation algorithms are available \cite{1992:Hwang}. In particular, polynomial time exact algorithms are available for some special cases, such as $|\mathcal{D}|=1,2,|\mathcal{I}|$. In fact, the algorithms proposed in the following subsections can also be used to solve the MST problem. The procedure for optimal CTI protection method is presented in Algorithm $2$.

\begin{algorithm}
\small
 \SetAlgoLined
 \SetKwData{Left}{left}\SetKwData{This}{this}\SetKwData{Up}{up}
 \SetKwRepeat{doWhile}{do}{while}
 \SetKwFunction{Union}{Union}\SetKwFunction{FindCompress}{FindCompress}
 \SetKwInOut{Input}{input}\SetKwInOut{Output}{output}
 \Input{$\mathcal{I}, \mathcal{D}, R$, edge weight vector $\mathbf{\bar{w}}$.}
 \Output{the set of transmission lines for CTI protection}
      Construct a weighted undirected graph $G=(\mathcal{V},\mathcal{E})$ by removing unmeasured edges and assigning weights to all the remaining edges.\;
      Find a minimum Steiner tree $T\left(\mathcal{D}\right)$ that connects the reference vertex and all the vertices in $\mathcal{D}$\;
      Keep the line information covert for all edges in $T\left(\mathcal{D}\right)$.
 \caption{Procedure for minimum-cost CTI protection}
\end{algorithm}

\subsection{Mixed defending strategy}
As mentioned in the last subsection, pure CTI method cannot protect $\mathcal{P}_2$-type vertices from undetectable attacks. Besides, in practice, not all transmission line information can be kept covert, such as those without transformer taps, etc. To tackle these problems, we develop a mixed defending strategy that allows the system operator to use both CTI and secure meter measurements to achieve the protection objective. Specifically, the system operator can select from all the meters to secure only a subset of them, and from a set of candidate transmission lines $\mathcal{K}_0$ to keep a subset $\mathcal{K}\subseteq \mathcal{K}_0$ line information covert.

\begin{figure}
\centering
  \begin{center}
    \includegraphics[width=2.5in]{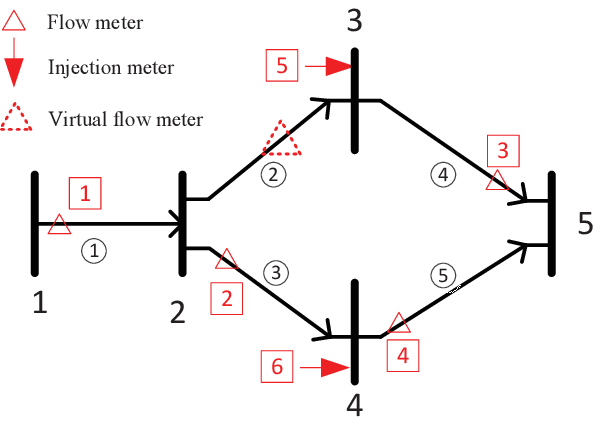}
  \end{center}
  \caption{Illustration of virtual power flow meter. The exact reactance of transmission line $e_2$ can be kept covert from attackers.}
  \label{54}
\end{figure}

Our first observation is that, it is useless to keep the line information of a bridging edge covert. In other words, bridging edges, if any, must be crossed out from $\mathcal{K}_0$. In fact, this is consistent with the result in the Steiner tree solution of pure CTI protection, where $\mathcal{K}_0$ is the set of all the measured edges. A Steiner tree containing a $\mathcal{P}_2$-type vertex and the reference node must contain at least one bridging edge. Therefore, after the bridging edges are crossed out from $\mathcal{K}_0$, there is no way to construct a Steiner tree solution. In other words, a $\mathcal{P}_2$-type vertex cannot be protected using the pure CTI protection method. Our second observation is that, from attackers' perspective, protecting a measured transmission line's information (a non-bridging edge of course) is equivalent to securing the line flow meter placed on it. If line flow meter is absent on the line, protecting the line information is as if ``installing" an extra secure flow meter on it that measures an arbitrary power flow direction, referred to as a virtual meter. This is because, when the reactance of $e_i$ is kept covert from attackers, they must introduce the same error at both ends of $e_i$ to avoid triggering the alarm. The net power flow change in $e_i$ must be zero before and after attack. This is as if a secure flow measurement is placed on $e_i$, regardless of its physical presence. The idea of virtual flow meter is illustrated in Fig. \ref{54}, where a covert transmission line $e_2$ is equivalently converted to a virtual meter measures the power flow on $e_2$. Notice that the virtual meters are only for the convenience of security analysis but provide no actual measurement readings.

Suppose that the system operator has the ability to protect meter measurements $\mathcal{M}$, consisting of both the actual meter measurements, and those \emph{virtual} flow meters converted from the candidate covert transmission lines in $\mathcal{K}_0$ free of bridging edges. The cost of protecting each actual meter measurement in dollar, e.g. manpower cost or surveillance installation fees, is given. The cost of a virtual flow meter is the cost of keeping the corresponding transmission line information covert. When a virtual flow meter collides with an existing actual flow meter, we assign the minimum of the costs as the cost to secure the flow meter. Then, the optimal mixed defending strategy can be obtained by solving the equivalent secure meter selection problem in \cite{2011:Bi}, i.e.
\begin{equation}
\label{90}
\begin{aligned}
& \underset{\mathcal{P}\subseteq\mathcal{M}}{\text{minimize}} & & W = \sum_{j\in \mathcal{P}} w_j\\
& \text{subject to} & & rank\left(\mathbf{\hat{H}}_{\{\mathcal{P}\},*}\right)= rank \left(\mathbf{\hat{H}}_{\{\mathcal{P}\},\{\mathcal{I}\setminus \mathcal{D}\}}\right) + |\mathcal{D}|,\\
\end{aligned}
\end{equation}
where $w_j$ is the cost in dollars of securing the meter $j$. $\mathbf{\hat{H}}$ is the new reduced measurement Jacobian matrix incorporating the virtual flow meters. $\mathbf{\hat{H}}_{\{\mathcal{P}\},*}$ is submatrix of $\mathbf{\hat{H}}$ consisting of the rows associated with the secure meter set $\mathcal{P}$. The detailed procedure of the mixed defending strategy is given in Algorithm $3$. I

It is proved in \cite{2013:Bi} that (\ref{90}) is an \textit{NP-hard} problem. However, it only studied a special case with uniform cost $w_j=1$ for all the meters. Here, we extend the methods in \cite{2013:Bi} to solve the general case with non-uniform costs. In the following, we first propose a MILP formulation to solve (\ref{90}), which significantly reduces the complexity compared with that of enumeration-based methods by capturing the topological structure of the optimal solution. Then, we introduce a tree-pruning heuristic to obtain an approximate solution in polynomial time.

\begin{algorithm}
\small
 \SetAlgoLined
 \SetKwData{Left}{left}\SetKwData{This}{this}\SetKwData{Up}{up}
 \SetKwRepeat{doWhile}{do}{while}
 \SetKwFunction{Union}{Union}\SetKwFunction{FindCompress}{FindCompress}
 \SetKwInOut{Input}{input}\SetKwInOut{Output}{output}
 \Input{$\mathcal{I}, \mathcal{D}, \mathcal{M}$, $R$.}
 \Output{a set of meters and transmission lines for mixed defending strategy}
      Construct an undirected graph $G=(\mathcal{V},\mathcal{E})$ by removing unmeasured transmission lines. Remove bridging edges in $\mathcal{K}_0$\;
      Convert each measured line in $\mathcal{K}_0$ into a flow measurement of arbitrary direction and assign a new weight to the measurement\;
      Solve (\ref{90}) to obtain the optimal protected meter set $\mathcal{P}^*$ for $\mathcal{D}$\;
      Restore the obtained solution into either covert transmission lines $\mathcal{K}$ or secure meter measurements $\mathcal{P}$.
 \caption{Procedure for mixed defending strategy}
\end{algorithm}

\subsection{MILP formulation of mixed defending strategy}
Recall in Proposition $1$ that a power system is observable if and only if an EMST can be found. The measurements in an EMST constitute a set of \emph{basic measurements}, protecting of which can defend all the state variables from undetectable attacks \cite{2010:Bobba}. Similarly, it is shown in \cite{2013:Bi} that no undetectable attack can be performed to compromise a subset of state variables $\mathcal{D}$, if an \emph{edge-measured Steiner tree} can be found to connect all the vertices in $\mathcal{D}$. In particular, each edge of the Steiner tree is mapped to a flow or injection meter that takes its measurement. A subtle difference is that, if an edge is mapped to an injection meter, all the vertices measured by the meter must be included in the Steiner tree. In Fig. \ref{61}, for instance, if $e_4$ is selected and mapped to $r_5$, then both $v_2$ and $v_5$ (vertices measured by $r_5$) must be included in the final edge-measured Steiner tree solution. If we assign a positive weight to each edge according to the difficulty of protecting the corresponding meter it is mapped to, the problem becomes finding a Steiner tree with the minimum edge weight sum. Consider a digraph $\overrightarrow{G}=\left(\mathcal{V},\mathcal{A}\right)$ constructed by replacing each edge in $G=\left(\mathcal{V},\mathcal{E}\right)$ with two arcs in opposite directions, where each arc is assigned the same weight as the original edge. Finding a minimum edge-measured Steiner tree is equivalent to the following minimum arc-measured Steiner arborescence (MASA) problem.

\textbf{Definition $5$: (MASA problem)} Given a digraph $\overrightarrow{G}=\left(\mathcal{V},\mathcal{A}\right)$, find a Steiner arborescence $\overrightarrow{T}^*= \left(\mathcal{V}^*,\mathcal{A}^*\right)$ and a set of meters $\mathcal{P}^*\subseteq \mathcal{M}$ that satisfy the following conditions
\begin{enumerate}
  \item $\mathcal{V}^*$ is the set of all vertices measured by $\mathcal{P}^*$;
  \item $\mathcal{D}\subset \mathcal{V}^*$ and $R\in \mathcal{V}^*$;
  \item each arc in $\mathcal{A}^*$ is one-to-one mapped to a unique meter in $\mathcal{P}^*$ that takes its measurement,
\end{enumerate}
with the minimum total arc weight $\sum_{j\in \mathcal{P}^*} w_j$.

Interestingly, the MASA problem has an equivalent network flow characterization. We set $R$ as the root and allocate one unit of demand to each vertex in $\mathcal{D}$. Commodities are sent from the root to the vertices in $\mathcal{D}$ through some arcs. Then, the vertices in $\mathcal{D}$ are connected to $R$ via the used arcs if and only if all the demand is satisfied. When we require delivering the commodity with minimum cost, the used arcs will form a directed tree $\overrightarrow{T}^*$, i.e. a \emph{Steiner arborescence}. In our problem, each arc in $\overrightarrow{T}^*$ must be mapped to a unique meter that takes its measurement. If an arc is mapped to an injection meter, all the vertices measured by the meter must be included in $\overrightarrow{T}^*$, as if an extra demand is allocated at these vertices. To distinguish from the actual demand at $\mathcal{D}$, we refer to the extra demand induced by the use of injection meters as \emph{pseudo demand}. Then, a MMSA can be found if we can construct a $\overrightarrow{T}^*$ with proper measurement mapping and satisfies both the actual and pseudo demand with minimum cost.

A MILP formulation of the MASA problem is

\begin{subequations}
\label{27}
\begin{align}
& \underset{\mathbf{X},\mathbf{Y},\mathbf{Z}}{\text{min}} & & \sum_{\left(i,j\right)\in \mathcal{A}}\bar{w}_{ij}\left(x_{ij}-z_{ij}\right) + \sum_{\left(i,j\right)\in \mathcal{A}}\bar{w}_{i} z_{ij} \label{26}\\
&\text{s. t. }  & & x_{ij}\geq \frac{y_{ij}}{q},  \;\; \forall \left(i,j\right)\in \mathcal{A} \label{21}\\
&  & & \mathbf{1}_E(i,j) + z_{ij} + z_{ji}\geq x_{ij}, \; \forall \left(i,j\right)\in \mathcal{A} \label{22}\\
&  & & \sum_{\left(i,j\right)\in \mathcal{A}} z_{ij}\leq \mathbf{1}_V(i), \;\; \forall i\in \mathcal{V} \label{23}\\
&  & & z_{ij} + z_{ji} \leq x_{ij} + x_{ji}, \;\; \forall \left(i,j\right)\in \mathcal{A} \label{28}\\
&  & & \sum_{\left(i,j\right)\in \mathcal{A}}y_{ij}- \sum_{\left(j,k\right)\in \mathcal{A}}y_{jk}= d(j), \forall j\in \mathcal{V} \setminus R \label{24}\\
&  & & x_{ij},z_{ij}\in \left\{0,1\right\}, \ y_{ij}\geq 0, \forall (i,j)\in \mathcal{A}. \label{25}
\end{align}
\end{subequations}
Here, $x_{ij}$ is a binary variable with $x_{ij}=1$ indicating that the arc $(i,j)\in \mathcal{A}$ is included in $\overrightarrow{T}^*$ and $0$ otherwise. $y_{ij}$ denotes the total amount of commodity through $(i,j)$. $z_{ij}$ is a binary variable with $z_{ij}=1$ indicating that the injection meter at vertex $i$ is mapped to arc $\left(i,j\right)$ or $(j,i)$, and $0$ otherwise. $\bar{w}_{ij}$ and $\bar{w}_i$ are the costs of protecting the flow meter on edge $[i,j]$ and the injection meter on vertex $i$, respectively. $q$ is chosen as a large positive number such that $q > y_{ij}$ always holds. $\mathbf{1}_E(i,j)$ and $\mathbf{1}_V(i)$ are two binary indicator functions, where $\mathbf{1}_E(i,j)=1$ if a flow meter is available at edge $[i,j]$ and $\mathbf{1}_V(i)=1$ if an injection meter is available at $v_i$. $d(j)$ is the demand at vertex $j$, where
\begin{equation*}
d(j)= \begin{cases}
1+\sum_{\left(j,k\right)\in \mathcal{A}} z_{jk} + \sum_{\left[k,j\right]\in \mathcal{E}}\sum_{\left(k,s\right)\in \mathcal{A}} z_{ks} &   j\in \mathcal{D}\\
\sum_{\left(j,k\right)\in \mathcal{A}} z_{jk} + \sum_{\left[k,j\right]\in \mathcal{E}}\sum_{\left(k,s\right)\in \mathcal{A}} z_{ks}   &   j\notin \mathcal{D}.\\
\end{cases}
\end{equation*}
For $j\notin \mathcal{D}$, $d(j)$ is the total pseudo demand. Otherwise, one extra unit of actual demand is counted as well.

The two terms in (\ref{26}) correspond to the costs on protecting flow meters and injection meters, respectively. When $\bar{w}_{ij}=\bar{w}_i$, the objective reduces to $\min \sum_{\left(i,j\right)\in \mathcal{A}} x_{ij}$, i.e. finding a Steiner arborescence with the minimum number of arcs as in \cite{2013:Bi}. Constraint $(\ref{21})$ forces arc $(i,j)$ to be included in $\overrightarrow{T}^*$ if any commodity flow passes through $(i,j)$. Constraint (\ref{22}) and (\ref{23}) ensure that each arc $(i,j)$ included in $\overrightarrow{T}^*$ has at least one measurement assigned to it and each injection measurement can only be assigned to at most one arc. Constraint (\ref{28}) eliminates the possibility that two injection meters at $v_i$ and $v_j$ are mapped to the same arc $(i,j)$. The flow conservative constraint (\ref{24}), together with $(\ref{21})$, forces the selected arcs to form an arborescence rooted at the reference vertex and spanning all vertices with positive demand. Once the optimal solution to (\ref{27}) is obtained, we can restore the optimal solution to the mixed defending strategy by protecting
\begin{enumerate}
  \item injection meter on bus $i$ if $z_{ij}=1$, $\forall (i,j)\in \mathcal{A}$;
  \item the transmission line information or the actual flow meter that corresponds to the flow measurement on arc $\left(i,j\right)$, if $x_{ij}=1$ and $z_{ij}=z_{ji}=0$, $\forall \left(i,j\right)\in \mathcal{A}$. That is, the arcs in $\overrightarrow{T}^*$ not mapped to any injection meters.
\end{enumerate}
Notice that the above MILP formulation can also be used to solve the MST problem in Algorithm $2$ by setting $z_{ij}=0$.

\subsection{Tree pruning heuristic}
Experimental results show that the MILP formulation in (\ref{27}) largely reduces the computational complexity compared with enumeration-based algorithms. However, solving a MILP is still \emph{NP-hard}, which may incur high computational complexity in large-scale power networks. To tackle the intractability of the problem, we introduce a tree pruning heuristic (TPH) that obtains an approximate solution in polynomial time.

\begin{figure}
\centering
  \begin{center}
    \includegraphics[width=3in]{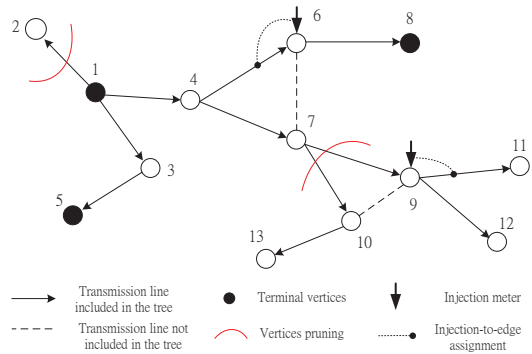}
  \end{center}
  \caption{An arc-measured arborescence. $\left\{v_1,v_5,v_8\right\}$ are the terminals and $v_1$ is the reference. Two marked edges ($[4,6]$ and $[9,11]$) are mapped to injection meters and the other unmarked edges are mapped to flow meters.}
  \label{63}
\end{figure}

The key observation is that, although it is generally hard to find a minimum arc-measured Steiner arborescence which connects a subset of vertices in the graph, it is relatively easy to find an arc-measured spanning arborescence that spans all the vertices. This can be achieved using the tree construction techniques proposed in \cite{1980:Krumpholz} or \cite{1986:Barglela}. Starting from an arborescence that spans all vertices in the full graph, our TPH method iteratively prunes away redundant vertices and updates the arborescence, until a shortest possible arc-measured arborescence (in the sense of total arc weight) is obtained. A pseudo-code is provided in Algorithm $4$. The TPH consists of multiple iterations of pruning operations. Each iteration consists of $3$ major steps:
\begin{enumerate}
  \item Arc-measured spanning arborescence generation (line $3-4$). We generate $K$ arc-measured spanning arborescences from the given vertices $\mathcal{\bar{V}}$ used for later pruning operation.
  \item Vertex identification and tree pruning (line $5-7$). In each of the $K$ arborescences, we first identify the sets of child vertices $C\left(i\right)$ and descendant vertices $D(i)$ for each vertex $v_i$. In particular, $v_j\in C\left(i\right)$ if there is an arc $(i,j)$ and $v_j\in D(i)$ if there exists a path from $v_i$ to $v_j$. In Fig. $\ref{63}$, for instance, $v_6$ and $v_7$ are the child vertices of $v_4$, while $v_6$ to $v_{13}$ are all descendent vertices of $v_4$. Starting from the root, we then find the largest prunable subset $C_s^*(i)$ for each $v_i$ in the arborescence. That is, the largest set of vertices in the descendent set of $v_i$, the pruning of which would still produce a residual arborescence as a feasible solution to the MASA problem.
  \item Vertex update (line $8$). We update $\mathcal{\bar{V}}$ from the best residual arborescence with the minimum arc-weight sum.
\end{enumerate}
The iteration continues until the arc weight of the best arborescence cannot be further reduced (line $9$). Then, the optimal mixed defending strategy can be easily obtained by protecting the injection meters in $\mathcal{P}^*$ and the flow meter (or the transmission line information) that corresponds to each flow meter in $\mathcal{P}^*$. The TPH method in this paper is a generalization of the pruning method in \cite{2013:Bi}, which considers all arcs with an equal weight. Due to the page limit, we refer interested readers to \cite{2013:Bi} for detailed tree pruning techniques. Instead, we provide an example in Fig. \ref{63} to illustrate its basic operations.

We consider an arc-measured arborescence with $12$ vertices. Starting from the root $v_1$, among the three child vertices of $v_1$, only $v_2$ can be pruned, since the descendent vertices of either $v_3$ or $v_4$ contain terminal vertex. After pruning $v_2$, we proceed to check $v_3$, whose only child vertex $v_5$ is a terminal. Then, we check $v_4$, where neither of its child vertices $v_6$ and $v_7$ can be pruned separately or together. On one hand, this is because $v_6$ contains terminal as its descendent vertices. On the other hand, the removal of $v_7$ does not remove the arc $\left(4,6\right)$, which is mapped to the injection meter at $v_6$ that measures $v_7$. For $v_7$, however, all of its descendent vertices can be pruned following the two pruning conditions. Up to now, we have finished the first iteration of pruning. Then, we use the remaining vertices $\left\{v_1,v_3,v_4,v_5,v_6,v_7,v_8\right\}$ to generate new arborescences, if any, and repeat the pruning iterations.

The purpose of introducing the parameter $K$ is because the final output $\mathcal{P}^*$ is closely related to the arborescence's topology obtained in Step $1$. Intuitively, with larger $K$, we have a larger chance to obtain an arc-measured arborescence with lower arc weight but also consume more computations. The proper choice of $K$ will be discussed in Simulations. The correctness of TPH is obvious from the following facts: $1$) the $K$ residual arborescences are always feasible to the MASA problem; $2$) the arc weight of the minimum residual arborescence is non-increasing during the iterations. There are at most $|\mathcal{I}|-|\mathcal{D}|$ rounds of pruning. In each round, $K$ arborescences are pruned and each takes $O\left(|\mathcal{I}|^3\right)$ time complexity, dominated by the Gauss-Jordan elimination computation. The overall time complexity is $O\left(K|\mathcal{I}|^4\right)$, which is considered efficient even for very large scale power systems.

\begin{algorithm}
\small
 \SetAlgoLined
 \SetKwData{Left}{left}\SetKwData{This}{this}\SetKwData{Up}{up}
 \SetKwRepeat{doWhile}{do}{while}
 \SetKwFunction{Union}{Union}\SetKwFunction{FindCompress}{FindCompress}
 \SetKwInOut{Input}{input}\SetKwInOut{Output}{output}
 \Input{$\bar{G}\left(\mathcal{M}\right)=\left(\mathcal{V},\mathcal{E}\right)$, $\mathcal{D}$, $R$, $K$ }
 \Output{Minimum protected measurements $\mathcal{P}^*$ to defend $\mathcal{D}$}
 \textbf{initialization:}  $\mathcal{\bar{V}}=\mathcal{V}$, $T^*=\emptyset$, $W^*=\infty$\;
      \Repeat{$W^* \geq \min \left(W_0,W_{-1}\right)$}{
      Record $T_{-1}=T^*$ and $W_{-1}=W^*$\;
      Find $K$ basic measurement sets of $\mathcal{\bar{V}}$, denoted by $\mathcal{\bar{P}}^k$, $k=1,..,K$. For each $\mathcal{\bar{P}}^k$, construct a feasible arc-measured spanning arborescences $T_k$. Let $W_0=\underset{k=1,...,K}{\min} \sum_{j\in \mathcal{\bar{P}}^k} w_j$ \;
      \For{\emph{each} $T_k$}{
            Starting from $R$ to all leaf vertices, find the largest prunable subset $C_s^*(i)$ for each $v_i$. Update $T_k=$ $T_k\setminus\left\{C^*_s(i)\cup D(C^*_s(i))\right\}$ until each vertex in $T_k$ is either processed or pruned\;
      }
      Select the trees $T^*$ with the minimum arc weight $W^*$ and update $\mathcal{\bar{V}}\leftarrow$ the vertices in $T^*$\;
}
 \eIf{$W_{-1}< W_0$}{
   $\mathcal{P}^*\leftarrow$ the remaining measurements corresponding to $T_{-1}$\;
   }
   {
   $\mathcal{P}^*\leftarrow$ the remaining measurements corresponding to $T^*$\;
   }
 \caption{Tree pruning heuristic algorithm}
\end{algorithm}

\subsection{Discussion of application scenarios}
The proposed defending mechanisms are designed to be applied to linearized DC power flow model. Essentially, however, the proposed mixed defending strategy is derived based on the general notion of \emph{topological observability} proposed in \cite{1980:Krumpholz}, which states that the observability of a power network, i.e. whether a unique estimate of all state variables can be derived from the meter measurements, is equivalent to whether it contains an arc-measured spanning arborescence that spans all the vertices in the graph. In this paper, we indeed generalize the conventional concept of network observability to state variable observability. That is, a unique estimate of a subset of state variables can be obtained from a subset of meter measurements. Similar to the arc-measured spanning arborescence structure for network observability, we propose an equivalent MASA structure for state variable observability, where a MASA connects the vertices of critical state variables to the reference bus. Based on the arborescence structure, we show that protecting a set of transmission line information and a set of meter measurements will render the attackers' effort to find a cut-set to compromise critical state variables impossible. Since the topological observability does not depend on power system line electrical parameters or operating points \cite{2013:Giani}, our results based on topological observability can also be applied to defend state estimation that uses nonlinear AC power flow models. Due to the scope of this paper, we do not extend the discussion and treat the detailed analysis in AC state estimation protection as a future working direction.

\begin{figure}
\centering
  \begin{center}
    \includegraphics[width=0.4\textwidth]{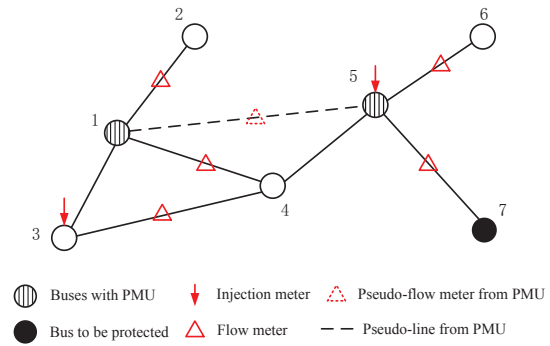}
  \end{center}
  \caption{Integration of PMUs in state estimation protection.}
  \label{pmu}
\end{figure}

Another interesting application scenario is to include phasor measurement units (PMUs) into the measurement set besides the power flow and injection measurements. Combined with GPS technology, PMUs can provide direct real-time voltage phasor measurement, i.e. voltage amplitude and phase angle, \footnote{There also exists other type of PMUs that can also provide current phasors of all the incident branches. We do not include them into consideration in this paper because they are inconsistent with our notion of a ``measurement", which provides only one reading at a time. However, we may study this problem in our future work.} with high precision and short measurement periodic time \cite{1996:Zivanovic}. Interestingly, our method can be easily extended to incorporate PMUs in the measurement set. Note that the state variable of a tagged bus cannot be compromised by attacks if a secure PMU is installed at the bus.\footnote{PMU is normally required to be installed at the reference bus to avoid the confusions due to the absolute voltage phasor measurements.} This is equivalent to installing a secure flow meter between the tagged bus and the reference bus. If there is a collide with existing flow meter, we merely change the cost of protecting the meter by the minimum cost between the existing flow meter and the PMU. On the other hand, if there exists no such power line connecting the two buses, a pseudo transmission line can be added to facilitate the calculation of the MASA problem. An illustrative example is given in Fig. $\ref{pmu}$, where a graph is extracted from a $7$-bus power network. Bus $1$ is the reference bus and PMUs are available at bus $1$ and $5$. The solid edges are the actual transmission lines in the power network. The dashed edge connecting bus $1$ and $5$ is made up by the PMU at bus $5$, where a pseudo-flow meter of random direction is placed on edge $e_{[1,5]}$. Since now we have formulated an equivalent problem with conventional power flows/injections measurements, the proposed arborescence construction algorithms can be directly applied. The only modification needed is that injection meters cannot be mapped to a dashed edge in the Steiner tree solution, because they do not measure the dashed edges in real system. The detailed modifications are omitted here to avoid the repetition of presentations. In the example in Fig. $\ref{pmu}$, an arc-measured arborescence can be constructed by edges $\left\{e_{[1,5]},e_{[5,7]}\right\}$, which are mapped to the pseudo-flow meter on edge $e_{[1,5]}$ (from the PMU at bus $5$) and the flow meter on edge $e_{[5,7]}$, respectively. Then, state variable of bus $7$ can be defended if the PMU at bus $5$ and the flow meter on $e_{[5,7]}$ are protected.

Before leaving this session, we want to specify the best use of the proposed algorithms. When protecting all the state variables, the state estimation protection problem in \cite{2010:Bobba} is a special case of ours. The proposed TPH algorithm indeed uses the same Gauss-Jordan elimination technique proposed in \cite{2010:Bobba}. For the proposed MILP formulation, however, the complexity could be much higher due to the NP-harness of solving integer programming problems. Therefore, we do not recommend to using MILP to solve the special case that all the state variables are to be protected. Another point to mention is the impact of the redundancy in measurements. On one hand, the complexity of the MILP increases with the measurement redundancy, as the number of variables $z_{ij}$ in the optimization problem (\ref{27}) will increase. On the other hand, the proposed TPH is not sensitive to measurement redundancy, i.e. its complexity is $O(K|I|^4)$, independent of the number of measurements.

\begin{table}
\caption{Statistics of Different Power System Testcases}
\footnotesize
\begin{center}
\begin{tabular}{|c|c|c|c|}
 \hline
  No. of buses &   $14$-bus    & $57$-bus & $118$-bus \\ \hline
  No. of lines &   $20$        & $80$     & $186$  \\ \hline
  Total no. of measurements &   $20$  & $80$     & $180$ \\ \hline
  No. of inject measurements &   $8$  & $30$     & $70$ \\ \hline
  No. of flow measurements &   $12$   & $50$     & $110$ \\ \hline
  No. of unmeasured lines &   $1$   & $2$     & $7$ \\ \hline
\end{tabular}
\end{center}
\end{table}

\section{Simulation Results}
In this section, we use simulations to evaluate the proposed attacking/defending mechanisms. All the computations are solved in MATLAB on a computer with an Intel Core2 Duo $3.00$-GHz CPU and $4$ GB of memory. In particular, MatlabBGL package is used to solve some of the graphical problems\cite{2006:Gleich}, such as maximum-flow/min-cut calculation, etc. Besides, Gurobi is used to solve MILP problems \cite{2013:Gurobi}. The power systems we considered are IEEE $14$-bus, $57$-bus and $118$-bus testcases, whose topologies are obtained from MATPOWER \cite{2007:Zimmerman} and summarized in Table I. For illustration purpose, a measurements placement of the 14-bus system is plotted in Fig. $\ref{71}$. The measurement placements for $57$-bus and $118$-bus systems are omitted for the simplicity of expositions.

\subsection{Min-cut partial knowledge attack}
\begin{figure}
\centering
  \begin{center}
    \includegraphics[width=3.2in]{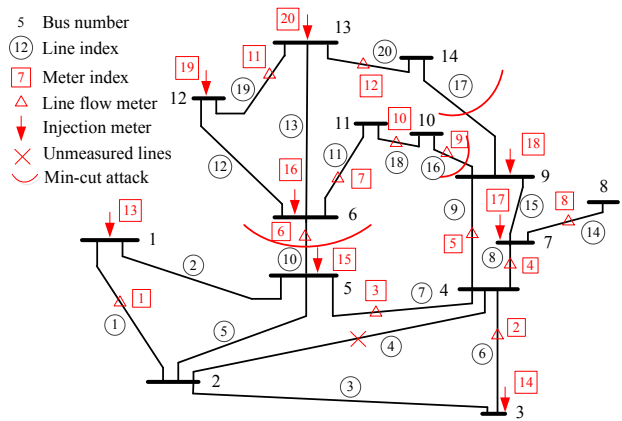}
  \end{center}
  \caption{A measurement placement of $14$-bus testecase. Notice that the unmeasured lines are those neither measured by a flow meter nor by any injection meter.}
  \label{71}
\end{figure}

In Fig. $\ref{71}$, we first illustrate the min-cut mechanism to formulate the optimal partial knowledge attack in the $14$-bus testcase, where the critical state variables to be protected are $\mathcal{D} = \left\{v_{10},v_{12}\right\}$ and $v_1$ is the reference. Without loss of generality, we assume that it takes the attacker $2$ dollars to obtain the exact reactance of lines $e_1$ to $e_9$, while $1$ dollar for lines $e_{10}$ to $e_{20}$. Following Algorithm $1$, a partial knowledge attack via min-cut calculation to compromise $\mathcal{D}$ is illustrated in Fig. $\ref{71}$. The edges on the optimal cut are $\left\{e_{10},e_{16},e_{17}\right\}$ and the minimum cost is $3$ dollars. Evidently, the optimal cut separates $\mathcal{D}$ from the reference bus $v_1$. Then, the attackers need to obtain the perfect knowledge of $\left\{y_{e_{10}},y_{e_{16}},y_{e_{17}}\right\}$ and inject adequate false data to meters $\left\{r_6,r_{9},r_{15},r_{16},r_{18}\right\}$ that measure the edges in the cut. For instance, false data $c_{10} y_{e_{16}}$ is injected to flow meter $r_9$ if the attackers intend to cause a bias $c_{10}$ to the estimate of $v_{10}$. The detailed injections to the other meters are omitted due to page limit. It can be easily verified using the residual test in (\ref{4}) that such an attack is undetectable.

\begin{figure}
\centering
  \begin{center}
    \includegraphics[width=2.2in]{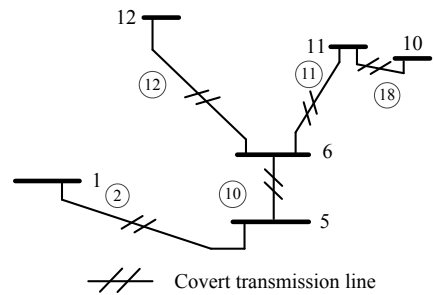}
  \end{center}
  \caption{The tree structure illustration of CTI protection method ($\mathcal{K}_0=\mathcal{E}$).}
  \label{72}
\end{figure}

\subsection{Case study of CTI protection}
\begin{figure}
\centering
  \begin{center}
    \includegraphics[width=2.2in]{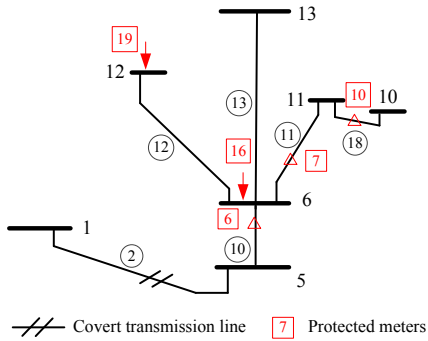}
  \end{center}
  \caption{The tree structure of mixed defending strategy ($\mathcal{K}_0=\left\{e_2,e_7,e_9,e_{16}\right\}$). The cost of protecting each transmission line is $1$ dollar. }
  \label{74}
\end{figure}

\begin{figure}
\centering
  \begin{center}
    \includegraphics[width=2.5in]{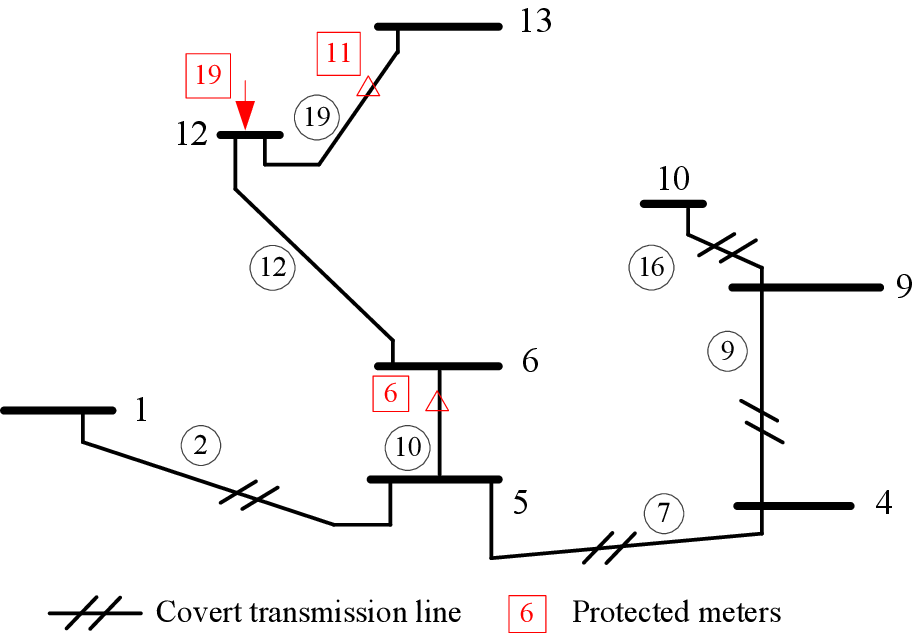}
  \end{center}
  \caption{The tree structure of mixed defending strategy ($\mathcal{K}_0=\left\{e_2,e_7,e_9,e_{16}\right\}$). The cost of protecting each transmission line is $0.1$ dollar.}
  \label{75}
\end{figure}

To better visualize the proposed CTI defending mechanisms, we present a case study for the $14$-bus system. The state variables to be protected are $\mathcal{D} = \left\{v_{10},v_{12}\right\}$. Following Algorithm $2$, we plot in Fig. $\ref{72}$ the Steiner tree solution of pure CTI protection, assuming that all the measured edges can be protected, i.e. $\mathcal{K}_0=\mathcal{E}$, and each costs $1$ dollar. We see that at least $5$ transmission lines should be kept from attackers. When only partial line information can be protected, e.g. $\mathcal{K}_0=\left\{e_2,e_7,e_9,e_{16}\right\}$, the CTI protection method would fail to defend $\mathcal{D}$ due to the insufficient resource needed for system security. In this case, we use the mixed defending strategy, where the cost of protecting each measurement is also assumed to be $1$ dollar. Following Algorithm $3$ (using the MILP formulation), the optimal defending strategy is illustrated in Fig. $\ref{74}$, where $\mathcal{K}^*=\left\{e_2\right\}$ and $\mathcal{P}^*=\left\{r_6,r_7,r_{10},r_{16},r_{19}\right\}$ and has a total cost of $6$ dollars. In practice, the cost of protecting line information can be much smaller than physically securing a meter measurement. Without loss of generality, we change the cost of protecting each line to $0.1$ dollar. As illustrated in Fig. $\ref{75}$, the optimal mixed defending strategy now becomes: $\mathcal{K}^*=\left\{e_2,e_7,e_9,e_{16}\right\}$ and $\mathcal{P}^*=\left\{r_6,r_{11},r_{19}\right\}$, with a total cost $3.4$ dollars.

\subsection{Efficiency of the MILP formulation}
We proceed to evaluate the computational complexity of the proposed MILP formulation. Extensive simulations show that the proposed MILP always obtain the optimal solution. Here, we use enumeration based (ENUM) exhaustive searching algorithm as the performance benchmark. For simplicity, we assume that $w_i=1$ in (\ref{90}) for each meter $i$. For MILP, we record the number of simplex iterations performed by the branch-and-cut algorithm ($I_M$). For the ENUM algorithm, we record the number of enumerations taken to obtain the optimal solution ($I_E$). Both numbers are the iterations consumed by the two methods to obtain a solution. Besides, we also record the CPU time of both MILP and ENUM algorithms, denoted by $T_M$ and $T_E$, respectively. The results in Table II are the average performance of $50$ independent experiments in $14$-bus, $57$-bus and $118$-bus systems. Without loss of generality, we randomly generate a $\mathcal{D}$ with size $|\mathcal{D}|=4$ in each experiment. We notice that both methods consume very similar CPU time to obtain an optimal solution in the $14$-bus system. However, the disparity becomes more and more significant as the network size increases. For instance, the MILP consumes $0.23$ seconds to obtain an optimal solution in the $57$-bus system, while the ENUM algorithm may take decades. For $118$-bus system, the CPU time of the ENUM algorithm can be practically considered as infinite, while the MILP formulation can solve the problem in around a minute. Similar results are also observed for the iteration numbers, i.e. $I_M$ and $I_E$, where ENUM becomes computationally infeasible even under moderate network size. The MILP methods largely reduces the complexity by exploiting the topological structure of the optimal solution. However, due to the \textit{NP-harness} of problem (\ref{90}), we still observe the fast growing complexity of MILP as the network size increases. For instance, the projected CPU time of MILP in a $300$-bus system is around $2$ days.

\begin{table}
\footnotesize
\caption{Comparison of MILP with enumeration based algorithm}
\begin{center}
\begin{tabular}{|c|c|c|c|c|c|c|c|}
 \hline
                            &       $T_M$     &  $T_E$           &   $I_M$          &  $I_E$       \\ \hline
  $14$-bus                  &   $0.041$ s     & $0.042$ s       & $69.3$       &   $250.0$    \\ \hline
  $57$-bus                  &   $0.23$ s      & projected 90 years         & $1354$       &   $6.79\cdot 10^{12}$     \\ \hline
  $118$-bus                 &   $64.6$ s        & $\infty$      & $874000$       &   $1.26\cdot 10^{21}$     \\ \hline
\end{tabular}
\end{center}
\end{table}

\subsection{Performance of the tree pruning heuristic}
We then evaluate the performance of the proposed TPH algorithm in both complexity and solution quality. The $14$-bus, $57$-bus and $118$-bus systems are considered and MILP is the benchmark for comparison. In each system, we assume that $20\%$ of the transmission lines can be chosen as the candidate covert transmission lines for protection. For instance, we consider $16$ out of $80$ transmission lines as $\mathcal{K}_0$ in the $57$-bus system. The cost of protecting each transmission line is $0.1$, while $1$ for securing a meter measurement. For TPH, we set the parameter $K=1$ and record the total number of vertices that are checked to produce a solution. For MILP, we record the number of simplex iterations performed by the branch-and-cut algorithm. Besides, we also record the CPU time for both methods.

\begin{figure}
\centering
  \begin{center}
    \includegraphics[width=3.2in]{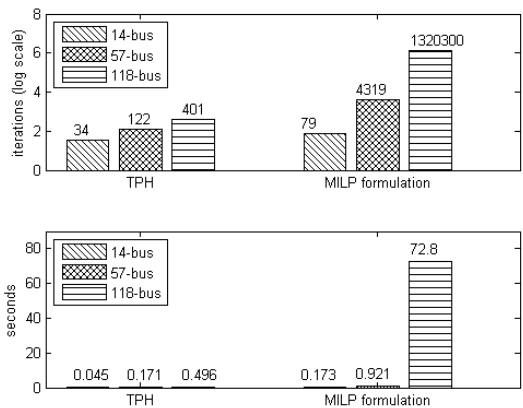}
  \end{center}
  \caption{Comparison of computational complexity for MILP and TPH. (a) The figure above shows the average number of iterations to obtain a solution; (b) the figure below shows the average CPU time to obtain a solution. }
  \label{66}
\end{figure}

Fig. $\ref{66}$ compares the computational complexity between MILP and TPH algorithms. The results in Fig. $\ref{66}$ are the average performance of $50$ independent experiments. Without loss of generality, we randomly generate a $\mathcal{D}$ with size $|\mathcal{D}|=4$ in each experiment. In Fig. $\ref{66}$a, we show the average number of iterations in log-scale for $14$-bus, $57$-bus and $118$-bus systems, respectively. The exact iteration numbers are also marked in the figure. We find that the iteration numbers are close for both methods in the $14$-bus system, where TPH consumes $34$ iterations and the MILP consumes $79$ iterations to obtain a solution. However, the difference becomes more and more significant as the network size increases. The number of iterations of TPH increases by $12$ times as the network size increases from $14$ to $118$ buses. In vivid contrast, the iteration number of MILP increases rapidly by $16713$ times, from merely $79$ to $1320300$. Similar results are also observed for the CPU time, where TPH takes only $0.496$ second to obtain a solution in $118$-bus system, while MILP consumes more than a minute, which is $421$ times slower than in the $14$-bus system. It is foreseeable that the computational complexity of the MILP method will become extremely expensive as we further increase the network size. For instance, the projected CPU time of MILP to solve a problem in $300$-bus system is around $2$ days, while it takes TPH less than $2$ seconds.

\begin{figure}
\centering
  \begin{center}
    \includegraphics[width=3.2in]{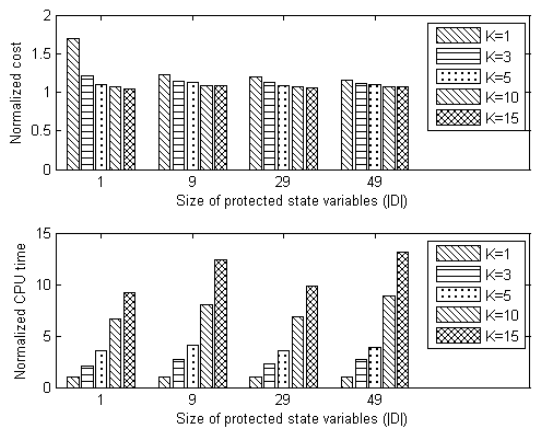}
  \end{center}
  \caption{Effect of $K$ to the performance of TPH in the $57$-bus system. (a) The figure above shows the cost of solution using TPH normalized by the optimal cost obtained by MILP; (b) the figure below shows the CPU time of TPH normalized by the CPU time when $K=1$.}
  \label{67}
\end{figure}

We also investigate the impact of the parameter $K$ to the performance of TPH. By varying the values of $K$ and $|\mathcal{D}|$, we plot the ratio $W/W^*$ for some selected $|\mathcal{D}|$'s in Fig. $\ref{67}a$, where $W$ is the cost of the solution obtained by TPH and $W^*$ is the minimum cost obtained from MILP.  We notice that the ratio improves notably for small $|\mathcal{D}|$ as $K$ increases from $1$ to $15$. For instance, the ratio improves from $1.70$ to $1.05$ for $|\mathcal{D}|=1$. The improvement is especially notable when we change $K=1$ to $3$. However, the improvement becomes marginal as we further increase $K$, such as the case with $|\mathcal{D}|=49$, where the ratio only improves by $0.09$ from $K=1$ to $15$. We also plot in Fig. $\ref{67}b$ the CPU time normalized against the time consumed when $K=1$. We observe that the CPU time increases almost linearly with $K$, which matches our analysis in Section IV. Results in Fig. $\ref{67}$ indicate that we should select a proper $K$ to achieve a balance between the solution quality and computational complexity. In particular, a large $K$, such as $K=10$, should be used when $|\mathcal{D}|$ is small relative to $n$, i.e. $|\mathcal{D}|<0.1n$. Otherwise, a small $K$, such as $K=3$, should be used when $|\mathcal{D}|$ is relatively large.

\section{Conclusions}
In this paper, we investigated the defending mechanisms against false-data injection attack using covert topological information (CTI). We studied from both the attackers' and the system operator's perspective and characterized the optimal protection as a well-studied Steiner tree problem in a graph. We also proposed a mixed defending strategy that bridges the gap between CTI protection and the conventional wisdom of secure meter measurement protection method. Both exact solution and reduced complexity approximate algorithms are proposed. The advantageous performance of the proposed defending mechanisms are evaluated in IEEE standard power system testcases.

\begin{IEEEbiography}
[{\includegraphics[width=1in,height=1.25in,clip,keepaspectratio]{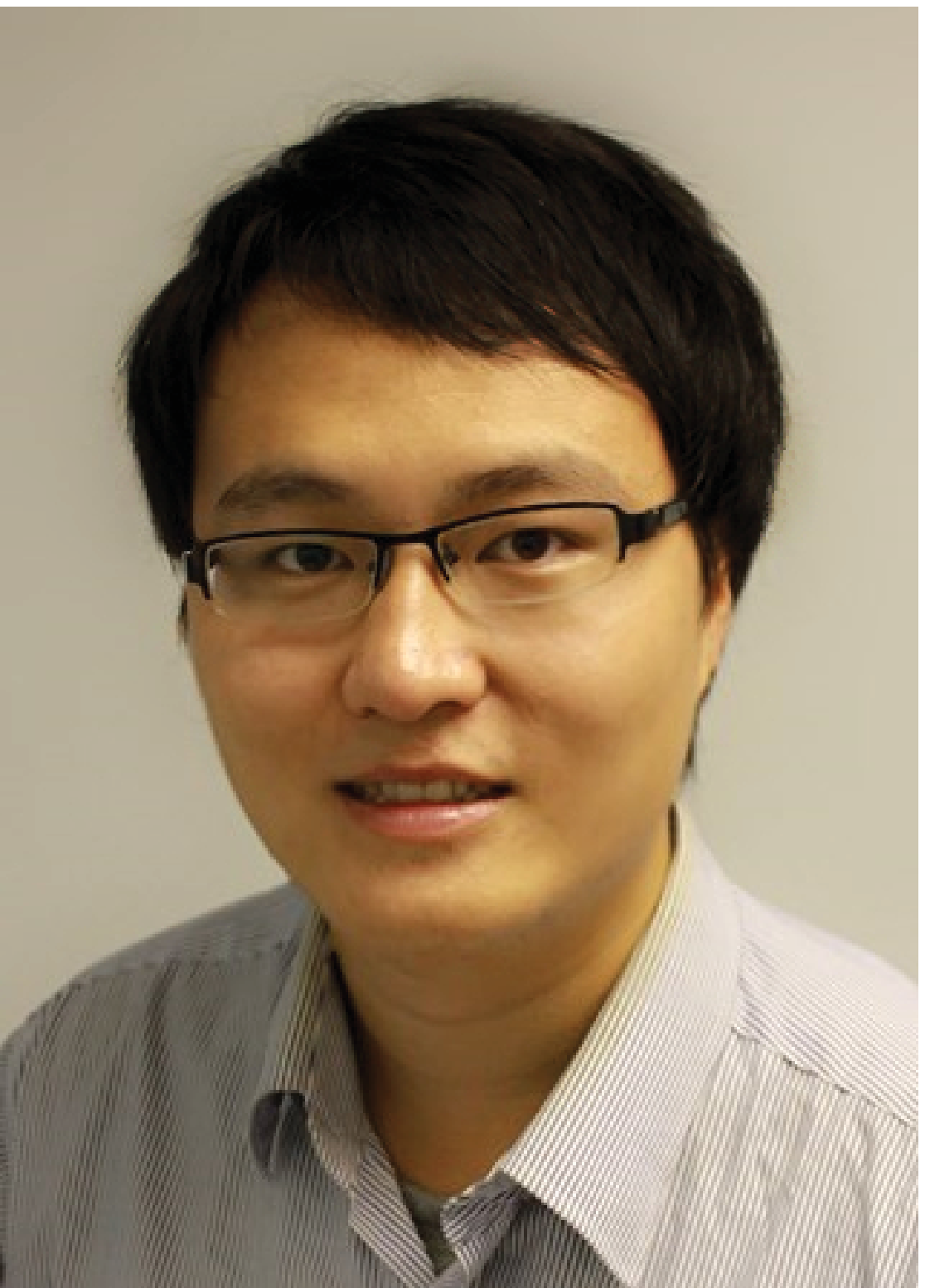}}]{Suzhi Bi}
(S'10-M'14) received his Ph.D. degree in Information Engineering from The Chinese University of Hong Kong, Hong Kong in 2013. He received the B.Eng. degree in communications engineering from Zhejiang University, Hangzhou, China, in 2009. He is currently a research fellow in the Department of Electrical and Computer Engineering, National University of Singapore, Singapore. From June to August 2010, he was a research engineer intern at Institute for Infocomm Research (I2R), Singapore. He was a visiting student in the EDGE lab of Princeton University in the summer of 2012. His current research interests include MIMO signal processing, wireless medium access control and smart power grid communications. He is a co-recipient of 2013 IEEE SmartGridComm Best Paper Award.
\end{IEEEbiography}

\begin{IEEEbiography}
[{\includegraphics[width=1in,height=1.25in,clip,keepaspectratio]{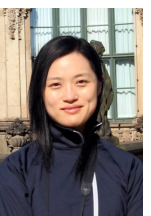}}]{Ying Jun (Angela) Zhang} (S'00-M'05-SM'11) received her Ph. D. degree in Electrical and Electronic Engineering from the Hong Kong University of Science and Technology, Hong Kong in 2004. She received a B. Eng in Electronic Engineering from Fudan University, Shanghai, China in 2000.

Since 2005, she has been with Department of Information Engineering, The Chinese University of Hong Kong, where she is currently an Associate Professor. She was with Wireless Communications and Network Science Laboratory at Massachusetts Institute of Technology (MIT) during the summers of 2007 and 2009. Her current research topics include resource allocation, convex and non-convex optimization for wireless systems, stochastic optimization, cognitive networks, MIMO systems, etc..

Prof. Zhang is an Executive Editor of IEEE Transactions on Wireless Communications and an Associate Editor of IEEE Transactions on Communications. She was an Associate Editor of Wiley Security and Communications Networks Journal and a Guest Editor of a Feature Topic in IEEE Communications Magazine. She has served as a Workshop Chair of IEEE ICCC 2013 and 2014, a TPC Vice-Chair of Wireless Communications Track of IEEE CCNC 2013, TPC Co-Chair of Wireless Communications Symposium of IEEE GLOBECOM 2012, Publication Chair of IEEE TTM 2011, TPC Co-Chair of Communication Theory Symposium of IEEE ICC 2009, Track Chair of ICCCN 2007, and Publicity Chair of IEEE MASS 2007. She was a Co-Chair of IEEE ComSoc Multimedia Communications Technical Committee, an IEEE Technical Activity Board GOLD Representative, 2008 IEEE GOLD Technical Conference Program Leader, IEEE Communication Society GOLD Coordinator, and a Member of IEEE Communication Society Member Relations Council (MRC). She is a co-recipient of 2013 IEEE SmartGridComm Best Paper Award, and 2011 IEEE Marconi Prize Paper Award on Wireless Communications. As the only winner from Engineering Science, she has won the Hong Kong Young Scientist Award 2006, conferred by the Hong Kong Institution of Science.

\end{IEEEbiography}

\end{document}